\documentclass[journal]{IEEEtran}

\pdfoutput=1

\usepackage[margin=1in]{geometry}
\usepackage{graphicx}
\usepackage{color}
\usepackage{subcaption}
\usepackage{float}
\usepackage{amsmath}
\usepackage{esint}
\usepackage{times}
\usepackage[latin9]{inputenc}
\usepackage{amsfonts}
\usepackage{amsthm}
\usepackage{amssymb}
\usepackage[nottoc]{tocbibind}
\usepackage{filecontents}
\usepackage{algorithm}
\usepackage[noend]{algpseudocode}
\usepackage{hyperref}
\usepackage{xcolor}
\usepackage{tabularx}
\usepackage{multirow}
\usepackage{array}
\usepackage{makecell}
\usepackage{wasysym}
\usepackage{bm}
\usepackage{colortbl} 
\usepackage{hhline}
\usepackage{color,soul}
\usepackage{algpseudocode}
\usepackage{tabularx,booktabs}
\usepackage{longtable}
\usepackage{multicol}

\newcolumntype{C}[1]{%
 >{\vbox to 11.5ex\bgroup\vfill\centering}%
 p{#1}%
 <{\egroup}}  

\usepackage{hyperref}
\hypersetup{
    colorlinks,
    citecolor=red,
    filecolor=black,
    linkcolor=red,
    urlcolor=blue
}
\hypersetup{linktocpage}

\numberwithin{equation}{section}
\numberwithin{figure}{section}

\begin{document}
\title{Empirical analysis of the variability in the flow-density relationship for smart motorways}

\author{Kieran Kalair and Colm Connaughton

\thanks{Manuscript received XX-XX-XX; revised XX-XX-XX; accepted XX-XX-XX. Date of publication XX-XX-XX; date of current version XX-XX-XX. This work was supported in part by the EPSRC and MRC Centre for Doctoral Training in Mathematics for Real World Systems under grant number EP/L015374/1.}%
\thanks{K. Kalair and C. Connaughton are based at the Centre for Complexity Science, University of Warwick, Coventry CV4 7AL, United Kingdom. }
}
\markboth{IEEE Transactions on Intelligent Transportation Systems,~Vol.~X, No.~X, MONTH~YEAR}%
{Kalair \MakeLowercase{\textit{}}:Empirical analysis of the variability in the flow-density relationship for smart motorways}

\maketitle

\begin{abstract}
The fundamental diagram is an assumed functional relationship between traffic flow and traffic density.
In practice, this relationship is noisy and exhibits significant statistical variability. 
On smart motorways, this variability is increased by variable speed limits that are not captured by the fundamental diagram. 
To study this variability, it is appropriate to consider the joint probability distribution function (pdf) of density and flow. 
We perform an empirical study of the variability in the relationship between flow and density using 74 days of data from 64 sections of London?s M25. 
The objectives are to determine how much of the variability in the flow-density relationship results from variable speed limits and to assess whether particular functional forms of the fundamental diagram are systematically preferred. 
Empirically, the joint pdf of flow and density is strongly bimodal, illustrating that traffic flows are often found in high-density or low-density regimes but rarely in between. 
We find that the high-density regime is strongly affected by variable speed limits whereas the low-density regime is not. 
The Daganzo-Newell (triangular) model of the fundamental diagram systematically fits best to the data. However, the optimal parameters vary with location. 
Clustering analysis of these parameters suggests three qualitatively different types of flow-density relationships applying to different sections of the M25. 
These clusters have natural interpretations in terms of the frequency and severity of flow breakdown. 
Accident rates also depend on cluster type suggesting possible links to other properties of traffic flows beyond the flow-density relationship. 
\end{abstract}


\linespread{1.0}

\section{Introduction}
\IEEEPARstart{T}he fundamental diagram of traffic flow is an important tool in understanding, categorizing and analyzing traffic flow. 
Despite several models and functional forms of the density-flow and density-speed relationships being suggested over the years, understanding the complex structure of fundamental diagrams is a difficult task.
Furthermore, analysis of real data suggests that the density-flow relationship in real traffic systems exhibits significant variability that is not captured by a single functional form.
As a result, it is natural to consider a distributional interpretation of the density-flow relationship, considering it as a 2-dimensional probability distribution and analyzing features of this to provide a different perspective on the relationship between density and flow.
In this work, we take such an approach to analyse the empirical variation in fundamental diagrams obtained from London's M25, one of the busiest Smart Motorways in the UK.
In addition to studying the variability of the density-flow relationship at a single location, we also consider variation between locations.
We approach this by first determining the optimal functional form for a diagram across 64 sections of road on the M25, then use clustering analysis to assess what qualitatively different types of density-flow relationships emerge in the network.
This provides insights into the variability in the density-flow relationship of smart-motorways on both a local and global scale.

\section{Literature Review}

Fundamental diagrams have been widely studied, starting with the first model of the density-speed relationship formulated by Greenshields in 1933 \cite{GreenshieldsFD}.
As an observation of a complex system, the diagrams contain much complexity, commonly exhibiting different levels of variation and spread of data as density varies.   
Many attempts to study fundamental diagrams have been conducted since their inception. 
One such work is \cite{Many_FundDiags_Paper_Wang}, where the authors took a year of data and proposed the optimal functional form of the fundamental diagram from this data was the logistic diagram.
Whilst most works take a piecewise triangular functional representation of a fundamental diagram, their search for, in their words, a functional form with `mathematical elegance and empirical accuracy' demonstrates that even a simple question of what curve fits a set of data points best is open to discussion.

Many sources of variation observed in the fundamental diagram are incorporated when combining the diagrams with models, either macroscopic or microscopic. 
One example of this is defining the flux function of a macroscopic model (which is simply the fundamental diagram) to have a random free-flow speed \cite{LWR_Subject_Uncertianty}. 
Whilst this captures some of the variation observed in real data and incorporates it in model results, it does not address the initial question of quantifying or better understanding the variation present in original data. 
Other works have argued that simply including random noise or terms in initially deterministic models can lead to behaviour inconsistent with the original model, which lead to \cite{Theoretical_Foundations_Stochastic} rephrasing this randomness in a Lagrangian framework model, essentially allowing driver headway choice to be stochastic.
Increasingly, instead of random flux functions, some modelling approaches use real-time data to update models which attempts to account for the deviations from typical behaviour, for example in \cite{Traffic_State_Data_Assim} and \cite{Traffic_State_Kalman_Filter}. 
Whilst useful, especially as more data and processing power becomes readily available, this again does not attempt to better understand the variation present before any modelling of traffic is completed.

Whilst our work focuses on fundamental diagrams of smart motorways, there is much work focusing on deriving them for an urban setting, for example \cite{Calibration_Probe_Vehicles} and \cite{GPS_Flow_Count_Conf}.
Whilst the first paper here attempts to create an algorithm to fit functions to the sparse probe data, and the second attempts to use the probe data to build a model of the traffic state, neither address the problem of variation in the urban sense.
Whilst discovering appropriate ways to use new data sources is important in the context of traffic management in urban scenarios, it is also important to better understand the situations where data collection is more routine and simple, for example on sensor equipped motorways.
It may be an interesting extension of our work to consider if there exist any significant differences in the urban level-variation compared to the more controlled and dense motorway traffic variation. 

Perhaps a work most similar to ours is \cite{VSL_Impact_On_Traffic}, where the authors investigate the impact of variable speed limits on the slope of a fundamental diagram, as well as the critical occupancy and capacity of a road. 
Our work differs substantially from this, as we take different interpretation of the fundamental diagram on a single link to analyse the impact of variable speed limits, through a 2d probability distribution and clustering approach, not a functional form approach. 
In addition, we consider the spatial consistency of different fundamental diagrams, something not covered in this paper

\section{Data Pre-Processing}\label{sec:Data}

The data for this study is taken from Highways England's National Traffic Information Service (NTIS). 
We choose to focus on the M25 motorway, one of the busiest in the country, taking 74 days of data starting from April 7th 2017. 

NTIS represents roads as a directed graph, with each motorway consisting of a number of `links' comprising the edges of the directed graph. 
On each link, the number of lanes is constant and no slip roads join or leave the motorway. 
We first recreate this directed graph model for the M25, consisting of 77 links, varying in length between 200 and 10,000 meters.

Placed on each link are physical sensors called `loops'. 
Loops detect passing vehicles at various distances along links, and from this the following measurements of interest are recorded:
\begin{itemize}
\item[-] Speed: The average speed of vehicles crossing the loop in kilometers per hour
\item[-] Flow: The number of vehicles crossing the loop in a time-period in vehicles per hour
\end{itemize}
We compute a time-series of density by dividing flow by speed.
NTIS aggregates this loop-level data to produce link-level recordings of speed and flow, among other quantities not used in this study.
The link-level data is the resolution we focus on here, meaning we extract one data-point measuring speed and flow every minute for 74 days, repeated for 77 different links.

In addition, we extract event and speed limit information. 
Events are any unusual activity on a link, separated into the following categories:
\begin{itemize}
\item[-] Accidents
\item[-] Vehicle obstructions
\item[-] General obstructions
\item[-] Abnormal traffic
\item[-] Maintenance, Road management \& poor environmental conditions 
\end{itemize}
We focus on accidents, vehicle obstructions, general obstructions and abnormal traffic events. 
NTIS defines an abnormal traffic event as a period when the true travel time on a link exceeds Highways England's predicted travel time by at least 120 seconds, for at least 5 consecutive minutes. The algorithm used to compute travel times is not publicly known.

Speed limit information represents operator interventions on motorway traffic. 
On some links on the M25, adjustable speed limit signs are displayed overhead to drivers, which can be used to temporarily change the speed limit on a section of road to any of 40, 50, 60 or 70 miles per hour.  

After processing, we have a set of time-series, with flags at each time there was an event or speed limit intervention on a link. 
Of the 77 links, 64 have working sensors and report data we can use for analysis.

\section{Exploratory Analysis}

\begin{figure}[ht!]
	\begin{subfigure}{.47\textwidth}
		\includegraphics[width=\textwidth]{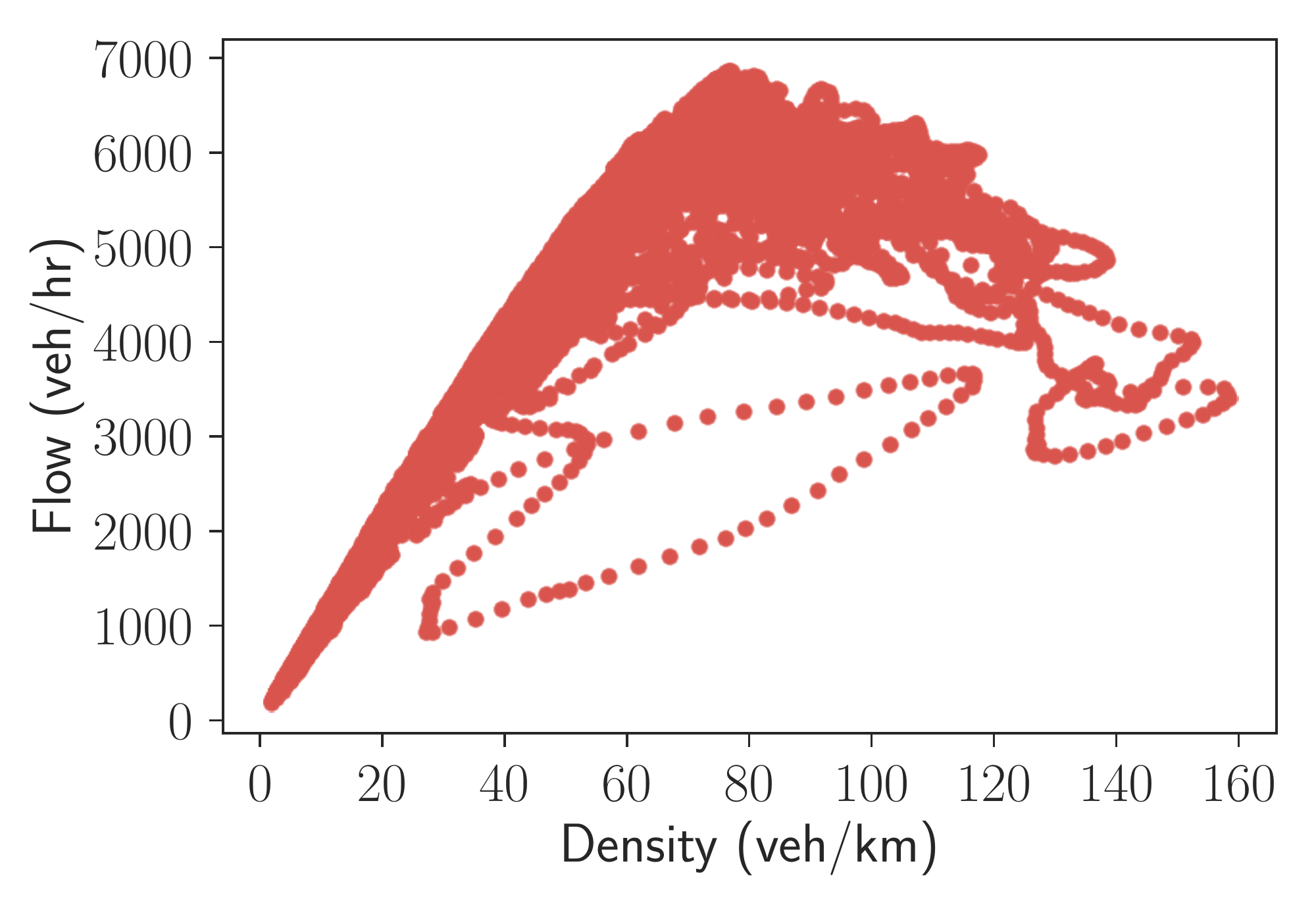}
	\end{subfigure}
	\caption[Example Fundamental Diagrams]{Example density-flow fundamental diagrams using 74 days of M25 data, taken from a link between junctions 2 and 3 (3636 metres long). }\label{fig:Example_FD_1}
\end{figure}

Typically, one views a fundamental diagram as a scatter plot of density and flow, as seen in Fig. \ref{fig:Example_FD_1}.

Inspecting Fig. \ref{fig:Example_FD_1}, we would consider the data to have quite a high amount of variation, particularly at density values around and above 60 veh/km. 
However, different methods of visualization offer different interpretations.
Suppose we now approximate the 2D probability distribution in \ref{fig:Example_FD_1} using kernel density estimation (KDE).
Kernel density estimation is a non-parametric method to estimate an underlying distribution function when only samples from it are available. 
For data points $\bm{X}_i \in \mathbb{R}^d$, $i \in \{ 1, 2, ..., N \}$ and in this context $d=2$, we write a kernel density estimate as: 
\begin{equation}
\hat{p}\left( \bm{x} \right) = \frac{1}{N}\sum_{i=i}^n k_{\bm{H}}\left( \bm{x} - \bm{X}_i \right)
\end{equation}
where $k_{\bm{H}}$ is some prior choice of kernel.
A common choice is the Gaussian kernel, which leads to the kernel density estimate being written as: 
\begin{equation}
\hat{p}\left( \bm{x} \right) = \frac{1}{N(2\pi)^{\frac{d}{2}}\left| \bm{\Sigma} \right|^{\frac{1}{2}} }\sum_{i=i}^n e^{ -\frac{1}{2}\left( \bm{x} - \bm{X}_i \right)' \bm{\Sigma}^{-1} \left( \bm{x} - \bm{X}_i \right) }.
\end{equation}

\begin{figure}[ht!]
	\begin{subfigure}{.47\textwidth}
		\includegraphics[width=\textwidth]{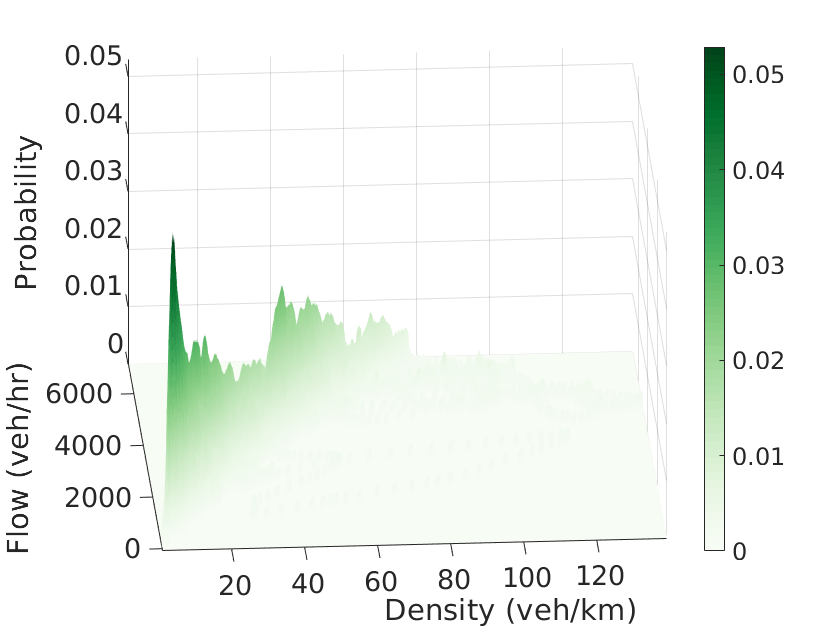}
	\end{subfigure}
	\caption[Example Fundamental Diagram (KDE)]{Example fundamental diagram using 74 days of M25 data, taken from a link between junctions 2 and 3 (3636 metres long). This is the same diagram as shown in Fig. \ref{fig:Example_FD_1}, however we take a vastly different interpretation of it. 
	From the original plot, we expect large variation, however from this plot we see the data is highly centered around 2 modes. Note the log-scale used on the coloring, further emphasising we have a very strong low-density peak, a less strong high density peak, and extremely rare excursions from these. }\label{fig:Example_FD_1_KDE}	
\end{figure}

The process of determining the optimal KDE is then to estimate the parameters of $\bm{\Sigma}$, which has 4 elements here.
There are various methods to do this, typically based on minimizing the mean integrated square error (MISE), defined as: 
\begin{equation}
MISE( \hat{p} ) = \mathbb{E}\left[ \int_{\mathbb{R}^d} \hat{p}(\bm{x}) - p(\bm{x}) d\bm{x} \right]
\end{equation}
with $p$ being some unknown true density function we wish to approximate.
Whilst the exact computation of MISE is not typically possible, approximation methods to it for multivariate data are given in \cite{multivariate_plug_in_bandwidth_selection_with_unconstrained_pilot_bandwidth_matrices}.
A KDE interpretation of a fundamental diagram is shown in Fig. \ref{fig:Example_FD_1_KDE}.

Inspecting Fig. \ref{fig:Example_FD_1} and Fig. \ref{fig:Example_FD_1_KDE}, we conclude that although significant deviations from typical behaviour exist, they are rare comparative to the overall behaviour observed on the link.  

\section{Analyzing the Single-Link Variation}

\begin{figure*}[ht!]
	\captionsetup{width=\textwidth}
	\begin{subfigure}{.48\textwidth}
		\includegraphics[width=\textwidth]{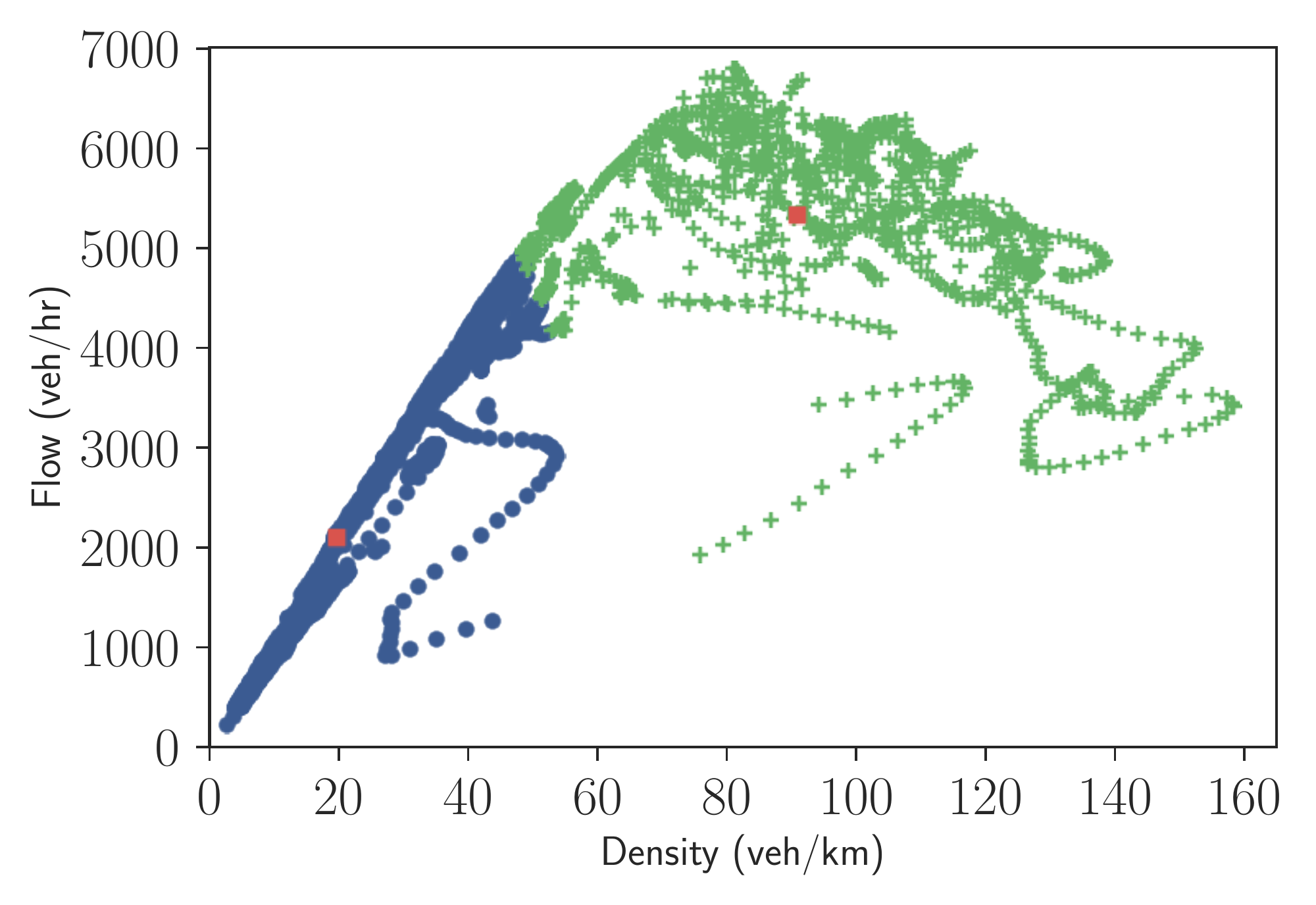}
		\caption{ 40 mph - Flow }
	\end{subfigure}
	\begin{subfigure}{.48\textwidth}
		\includegraphics[width=\textwidth]{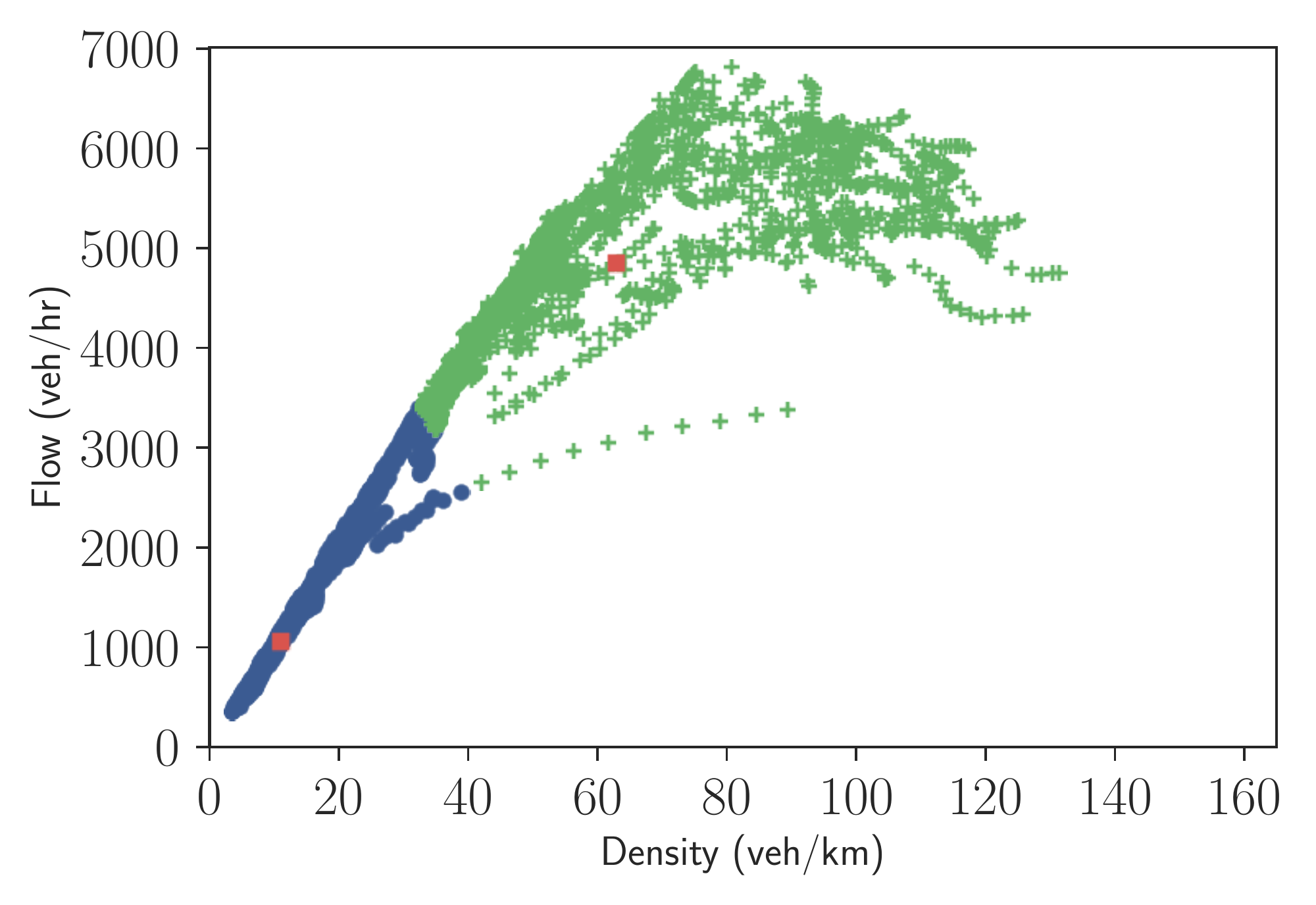}
			\caption{ 50 mph - Flow }
	\end{subfigure}
	\begin{subfigure}{.48\textwidth}
		\includegraphics[width=\textwidth]{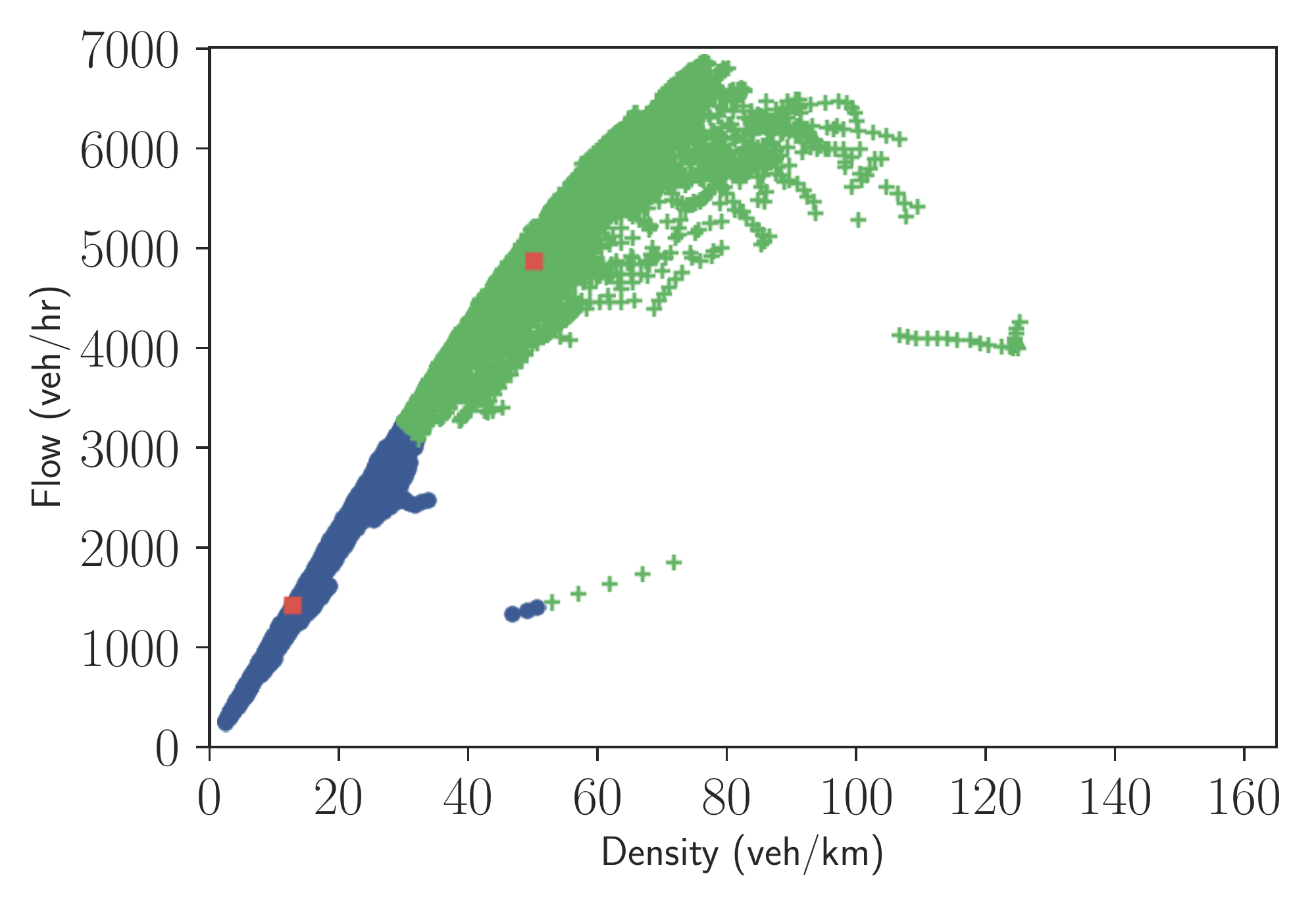}
			\caption{ 60 mph - Flow }
	\end{subfigure}
	\begin{subfigure}{.48\textwidth}
		\includegraphics[width=\textwidth]{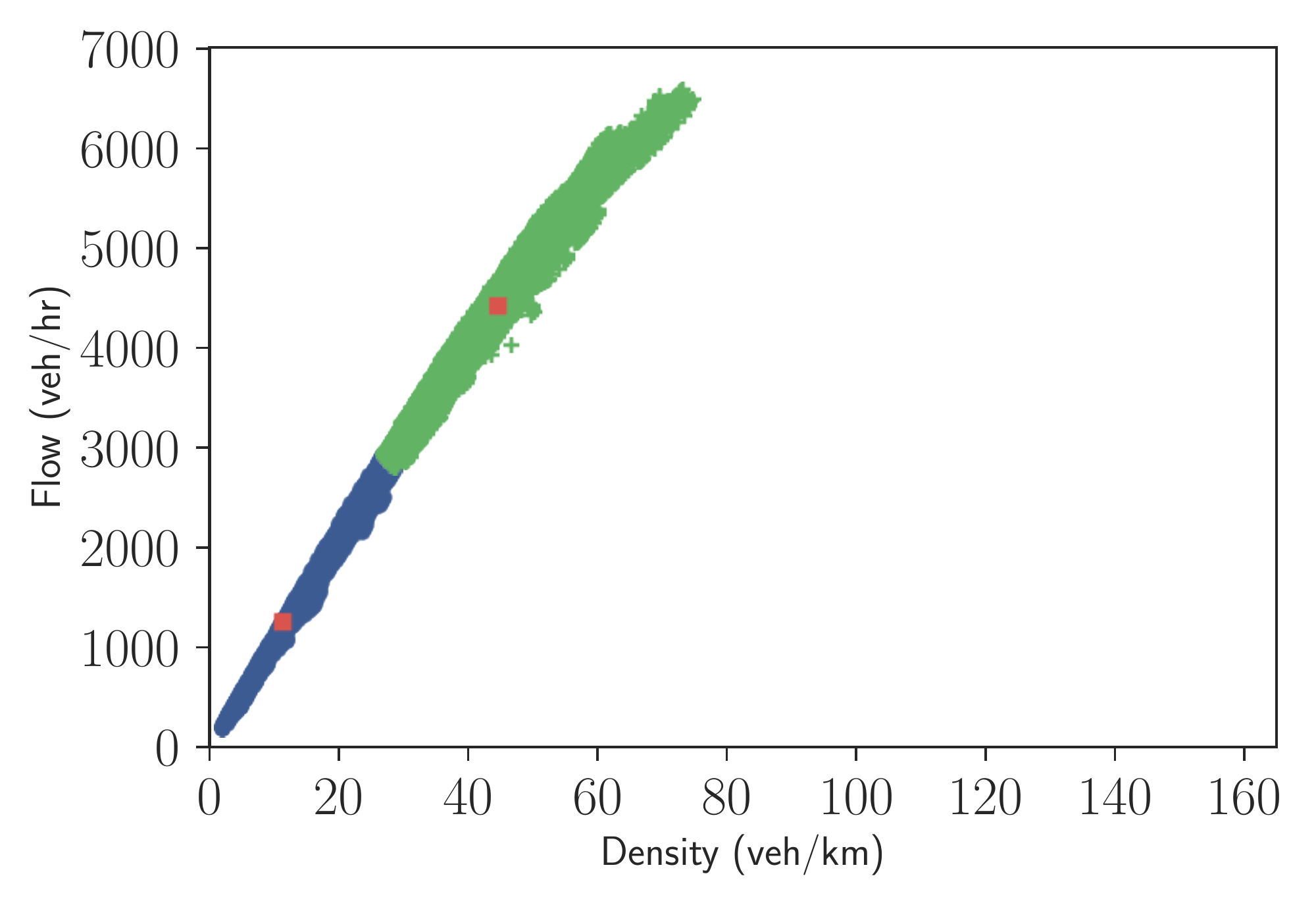}
		\caption{ 70 mph - Flow }
	\end{subfigure}

	\caption[Clustered Fundamental Diagram Segmented by Speed Limit]{ Example fundamental diagrams using 74 days of M25 data, taken from a link between junctions 2 and 3 (3636 metres long). The data has separated into separate cases, each having a different variable speed limit active. We see the diagram when the speed limit on the link is 70 miles per hour is not at all representative of the diagram when the speed limit is set to 40 mph. For reference, we have the conversions: (40, 50, 60, 70) mph being (64.4, 80.5, 96.6, 112.7) km/hr.
	The colouring and marking of points refers to the clustering analysis discussed in section \ref{subsec:SingleLinkClusteringAnalysis}.
	The medoids are shown in red ($\square$), with the points lying the in low-density mode shown by green $(+)$ and high density mode shown by blue $(\circ)$.
	}\label{fig:Clustered_FD_Exaxmple}
\end{figure*}

To begin the analysis in this section, suppose we first take a single link. For this link, we know when speed limits are on, so we segment our series into these categories.
In particular, as speed limits are inherently linked to events (and may be activated by operators when an event seems likely, or has occurred and traffic is disrupted), we can consider how much of the variation seen in each fundamental diagram can be explained by the variable speed limit active.
Any link on the M25 may contain some number of speed limit signs. 
At any point where we have a set of signs, we will have one above each lane, currently all displaying the same speed limit for each lane. 
However, different groups of signs on a link may have different speed limits active, meaning there is some ambiguity as to what we specify the current speed limit on a link to be in these cases.
As a result, we group the diagrams as follows: when we detect at-least one variable speed limit sign has been altered on a link, we calculate the average speed limit displayed on all of the changed signs, and set the current speed limit on a link to the nearest value in the feasible speed limit set to this average.
The feasible speed limit set is given in miles per hour as (40, 50, 60, 70), and in kilometers per hour as (64.4, 80.5, 96.6, 112.7)
Results for segmenting a diagram into different speed limit cases are shown in Fig. \ref{fig:Clustered_FD_Exaxmple}.
 
Inspecting Fig. \ref{fig:Clustered_FD_Exaxmple}, we see two significant behaviours.
Firstly, there are no data points past densities around 80 vehicles per kilometer in the 70 mph set. 
Rather than this being a setting rule, it is likely that this is simply a consequence of some other setting rule, as density is not a quantity directly provided by NTIS data.    
Secondly, it is striking how the variation in the data, particularly flow, decreases as we increase the speed limit.
Clearly, 40 mph speed limits will be active in the most extreme cases of events and traffic state, and 70 mph speed limits will be active in the most ideal of scenarios.

Whilst the speed diagrams still see vast decreases in variation as we increase the speed limit from 40mph to 70mph, they are perhaps not so extreme as in the flow case. 
It should also be noted that, although it appears many data points lie above the speed limit in each diagram, there are multiple reasons for this. 
Apart from drivers simply ignoring speed limits, it is likely that when only part of a section of road has a particular lower speed limit on, other sections may be higher, meaning the average speed measurement we receive from NTIS may be higher than the minimum speed limit.
Also, in the case of 40 mph speed limits, these are often set for maintenance work late at night, so the opportunity for drivers to exceed them is far greater than say during an accident event, where a physical queue forces drivers to slow down.

\subsection{Single Link Clustering Analysis}\label{subsec:SingleLinkClusteringAnalysis}

Having identified this clear distinct behaviour for each speed limit case, we now try to quantify it. 
Through our various methods of exploratory analysis on the fundamental diagrams, we saw in Fig. \ref{fig:Example_FD_1_KDE} that the distribution of data in the fundamental diagrams was  bi-modal.  
As a result, it is natural to consider measuring variation in the entire diagram relative to each mode. 
If some other measure, for example the 2-d analogy to standard deviation was used, we would neglect this clear behaviour present in our data. 
First, we must identify the locations of the modes, then come up with a measure of distance around them, finally comparing the spread around these modes for data from each speed limit case. 

It should be noted that we could simply take the two maxima of our KDE fit to determine the mode locations, however using clustering ensures we have no approximation errors and the tails are correctly accounted for.

To determine the location of the modes, clustering is insightful and practical technique.
In particular, as we are aware a-priori that our data is concentrated in two locations, we use clustering algorithms to determine these locations.
There are various different clustering algorithms one could use, however we have a significantly simplified selection procure as we are searching for exactly 2 clusters in each diagram.
Classic choices, based on the prior need for 2 clusters, are k-means and k-medoids. 
Whilst widely applied to various problems, k-means can be significantly influenced by outliers in the data, which are clearly present here. 
Instead, we can use k-medoids, which attempts to assign cluster centers that are themselves data-points, rather than averages of the points. 
It is widely known that k-medoids is a prohibitively expensive algorithm when using a naive implementation. 
Instead, we use the `clara' variation \cite{CLARA_Clustering}, clustering subsets of the data of a fixed size, and running over 50 random starting conditions to attempt to avoid initial condition bias.  
Before clustering, we scale the data to ensure all quantities have similar ranges, dividing the speed values by the maximum observed speed, flow by the maximum observed flow, and density by the critical density at which the maximum flow occurs. 

Clustering provides approximate mode locations for each diagram, with an example shown in Fig. \ref{fig:Clustered_FD_Exaxmple}.
From Fig. \ref{fig:Clustered_FD_Exaxmple}, it is clear that for all speed limit cases, the low density mode stays roughly constant in position, however the high-density model moves significantly. 
For the flow cases, we see it moves down in density as the speed limit increases, and slightly down in flow. 
The speed cases are similar, with the high density mode moving up in speed and down in density as speed limits increase, however one should also note the significant reduction in spread around the modes as a function of which speed limit is active. 

\begin{figure}[ht!]
	\centering
	\includegraphics[width=\linewidth]{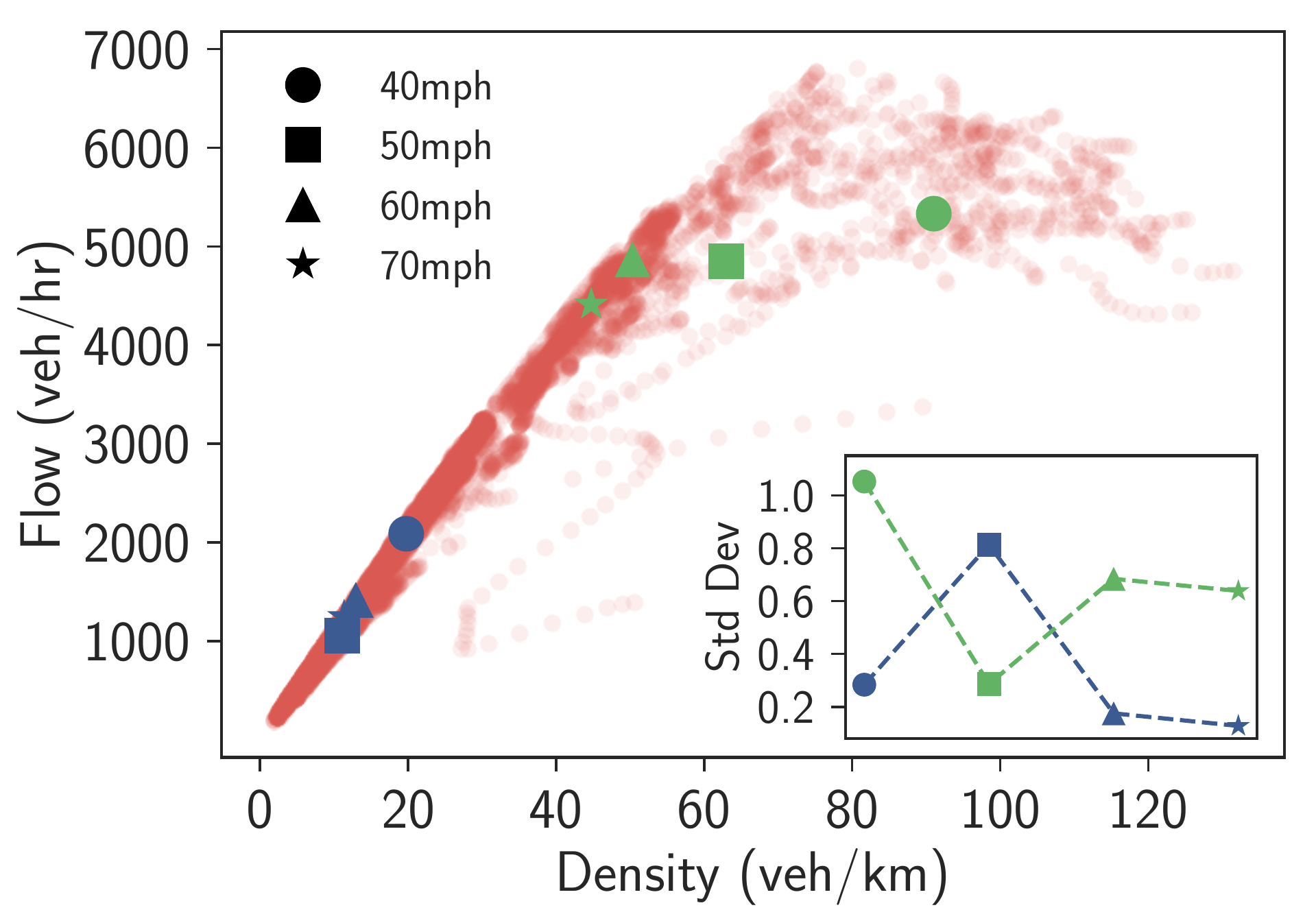}
	\caption[Clustering Mode Analysis]{ A fundamental diagram, with the high and low density clustering modes at each speed limit value highlighted. Notice how the movement of the high-density mode is far greater than the movement of the low-density mode, which essentially remains in the same position when speed limits greater than 40mph are active.
	In the insert, we plot the standard deviation of each cluster for each speed limit value.
	We see when speed limits are set to 50mph, the low-density mode has larger standard deviation, however at all other values the high-density mode has higher standard deviation.
	}\label{fig:ModeLocationVsSpeedLimit}
\end{figure}

To quantify this behaviour, we consider the location of each mode as a function of speed limit, with results shown in Fig. \ref{fig:ModeLocationVsSpeedLimit}.

From Fig. \ref{fig:ModeLocationVsSpeedLimit}, we see clear verification of the visual conclusions previously made. 
Importantly, we see significant shifts in the density coordinate for the high density flow mode when compared to the low density case.  
Quantifying this, we see that the high-density mode moves from around 90 veh/km when 40mph speed limits are active to around 45 veh/km when 70mph speed limits are active, a relative change of $44\%$.
Comparatively, we see the low-density mode move from a density of around 20 veh/km when 40mph speed limits are active to a density of around 12 veh/km when 70mph speed limits are active, a similar relative change of $40\%$.
It should be noted however that most of the low-density mode movement occurs between the 40mph and 50mph cases, with its location essentially being fixed for the 50, 60 and 70mph cases. 
On the other hand, the medoid flow coordinates change from 5300 veh/hr to 4400 veh/hr in the high-density case and 2000 veh/hr to 1250 veh/hr in the low-density case.
Again, most of the low-density mode changes occur between the 40 and 50 mph cases.

Additionally, the insert in Fig. \ref{fig:ModeLocationVsSpeedLimit} allows us to compare the spread around each medoid for each of the speed limit values.
We see the standard deviation for the low-density mode actually peaks when 50mph speed limits are active, however decreases from the 70 to 50 to 40mph cases.
The peak is standard deviation of the low-density mode coincides with the lowest standard deviation of the high-density mode.
Recall that all quantities were scaled by characteristic values before fitting, and we are comparing the euclidean distances between points in each cluster and the cluster medoid, so this standard deviation is computed on dimensionless values.

Inspecting the distance distributions themselves, we generally see heavy-tailed distributions, but over very short ranges.
When considering appropriate fits to these distributions using the methods described in \cite{PowerLawsInEmpericalData}, the limited range of distances makes distinguishing between any heavy-tailed distributions difficult.

\section{Analyzing the Link-to-Link Variation}

\subsection{Fitting Functional Forms to Data}

\begin{figure}[ht!]
	\centering
	\begin{subfigure}{\linewidth}
		\includegraphics[width=\textwidth]{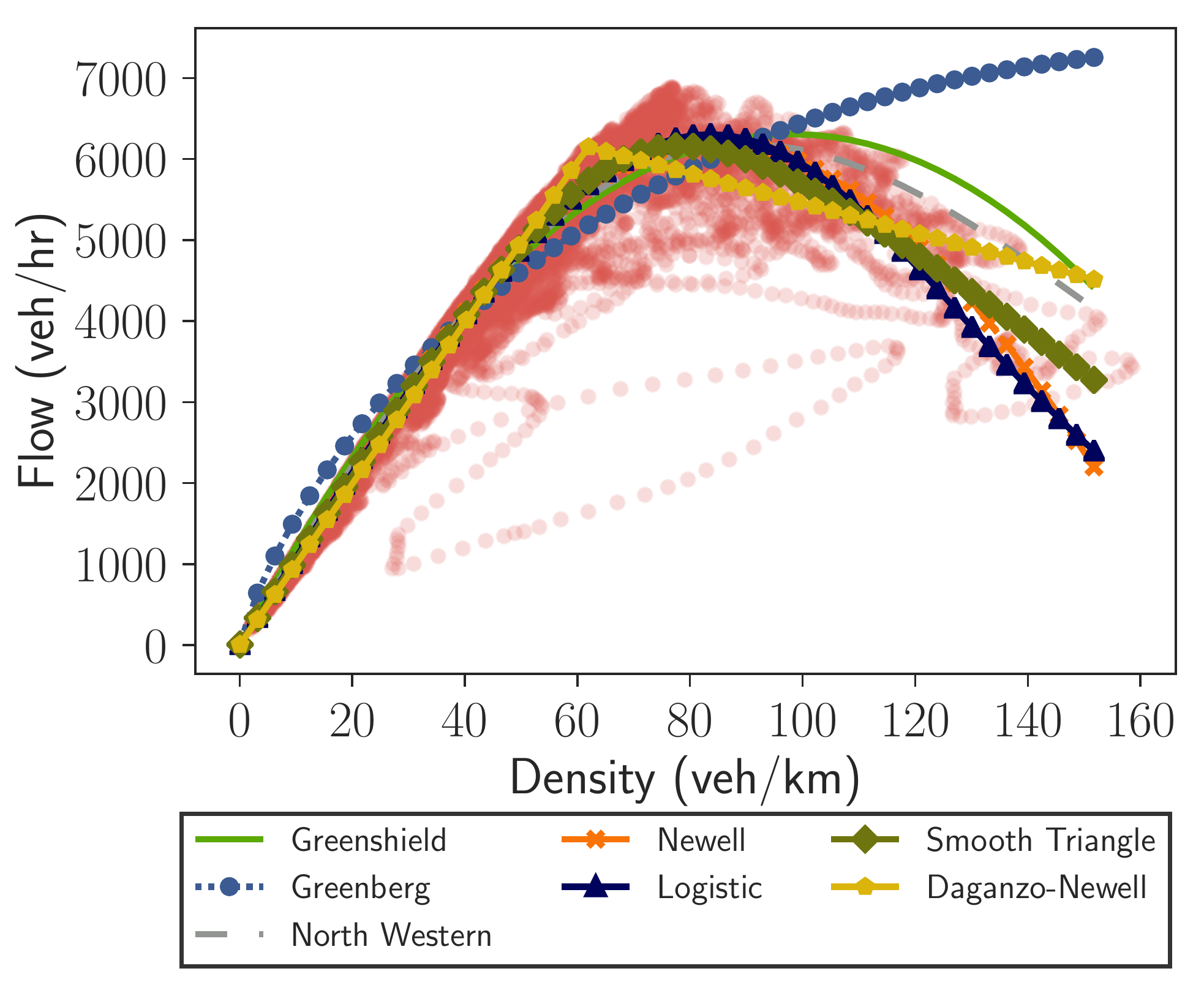}
		\caption{ Realistic Parameter Space }
	\end{subfigure}
	\caption[Fundamental Diagram Functional Fits]{ Results for fitting each functional form of fundamental diagram to a single links fundamental diagram. All inputs, limits and data are scaled before fitting. }\label{fig:FD_Functional_Fits}
\end{figure}

Having investigated the variation in the fundamental diagram of a single link, we now consider appropriate functional fits to the data, aiming to determine which of the many available functional forms best describes the data.
We outline the most common fits seen in the literature in Appendix \ref{appendix:FuncForms} table \ref{table:FundDiags}.
To begin, we simply fit each of the functional forms given in table \ref{table:FundDiags}, and show our results in Fig. \ref{fig:FD_Functional_Fits}.

\begin{figure}[ht!]
	\centering
	\includegraphics[width=\linewidth]{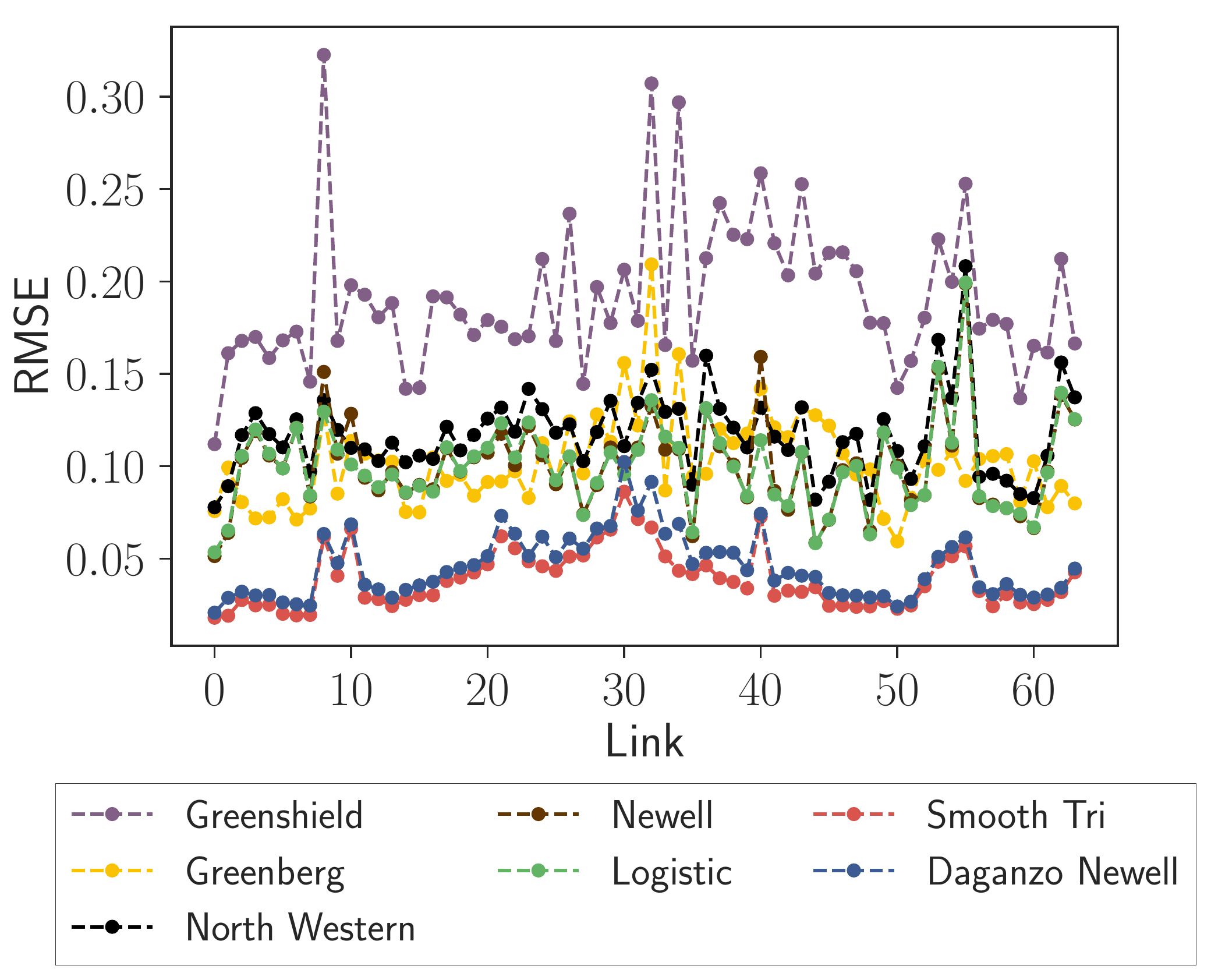}
	\caption[Optimal Diagram Analysis For All Links]{ Root mean square error results for all links considered. We see the Daganzo Newell and Smooth Triangle results are generally indistinguishable across links, however these two diagrams consistently have lower RMSE than the others considered.
	}\label{fig:DiagramRMSE_Vs_Link}
\end{figure}

Inspecting Fig. \ref{fig:FD_Functional_Fits}, we see generally similar fits to the data across all diagrams for low density values, with the Greenberg diagram clearly being the poorest fit at both low and high densities. 
Any diagram we use should be representative of all data however, not a single link.
To identify which diagram is best representative of the data, or indeed if such a diagram exists, we fit all of the considered diagrams to each of our 64 links, again running 100 different optimizations for each diagram on each link to attain the optimal parameter sets of each diagram.
In Fig. \ref{fig:DiagramRMSE_Vs_Link}, we show the results of measuring the root mean square error (RMSE) for all links, using only the optimal parameter set in each case.

From Fig. \ref{fig:DiagramRMSE_Vs_Link}, it is clear that the most representative functional form for the density-flow relationship is triangular, with the discrete and continuous versions consistently achieving lower RMSE than any other consider functional form. 
An interesting point to consider is that the Daganzo-Newell fundamental diagram peak (representing maximum flow) is not placed at the largest observed flow value. 
Doing so actually decreases the goodness of fit, leading to a smaller $R^2$ value and larger root-mean square error and sum of squared error. 
We instead capture the relationship with the majority of the data by placing it at a position much in agreement with Fig. \ref{fig:Example_FD_1_KDE}.
As the Daganzo-Newell triangular diagram is by far the most common in the literature, we show results for this from now on.

\subsection{Determining the Optimal Diagram During Speed Limit Categories}

Whilst we have seen that the optimal diagram, in the sense of minimum RMSE, for data from each of our links is triangular, it is of interest to investigate if the same is true when segmenting diagrams by speed limit.
To do so, we again fit each candidate diagram to all of our links, however this time fitting them to subsets of data, each containing points only when specific speed limits are active. 
We then attain goodness of fit results for each subset across all links, 61 of which have variable speed limit signs on.

\begin{figure}[ht!]
	\centering
	\begin{subfigure}{0.48\textwidth}
		\includegraphics[width=\linewidth]{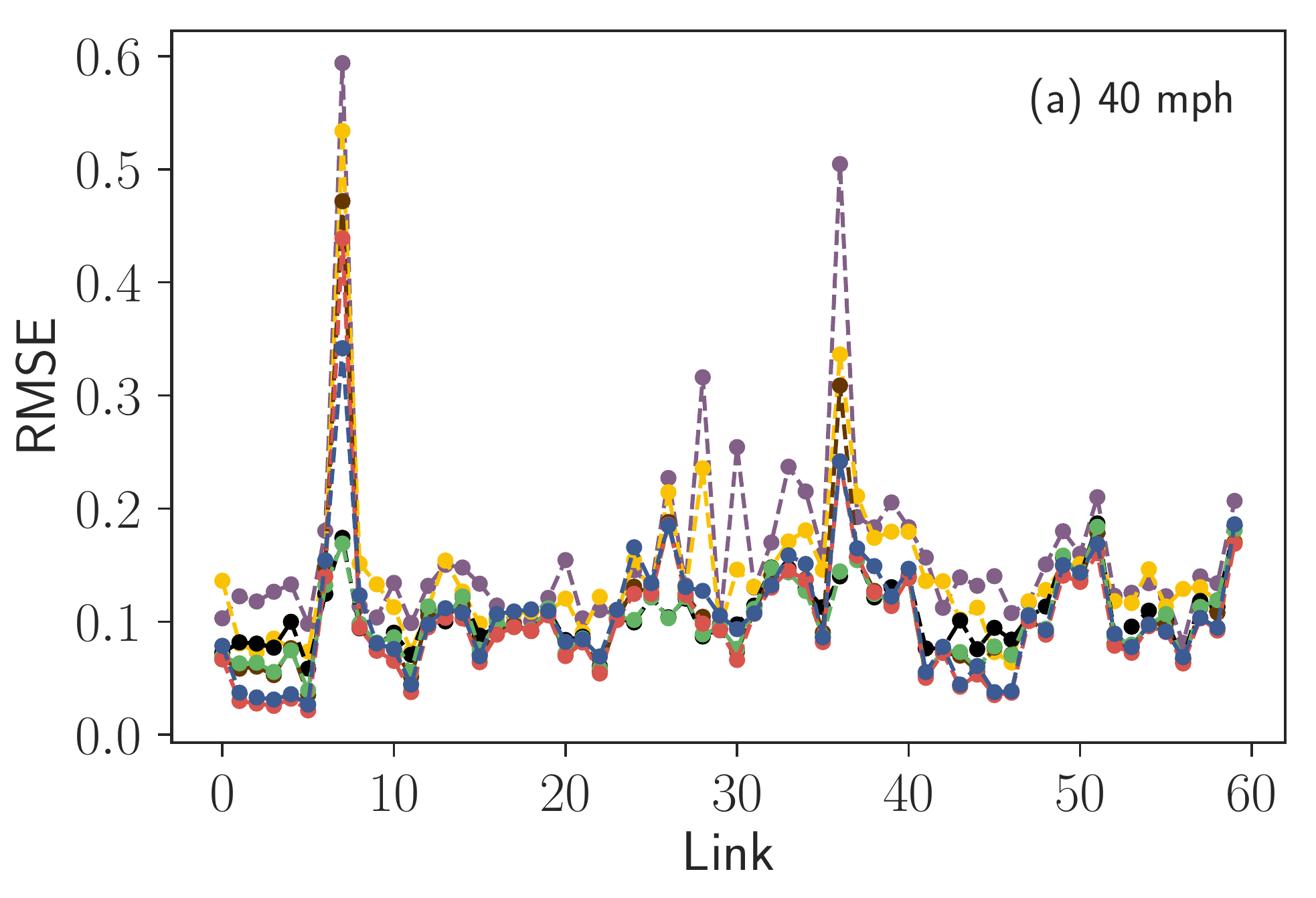}
	\end{subfigure}
	\begin{subfigure}{0.48\textwidth}
		\includegraphics[width=\linewidth]{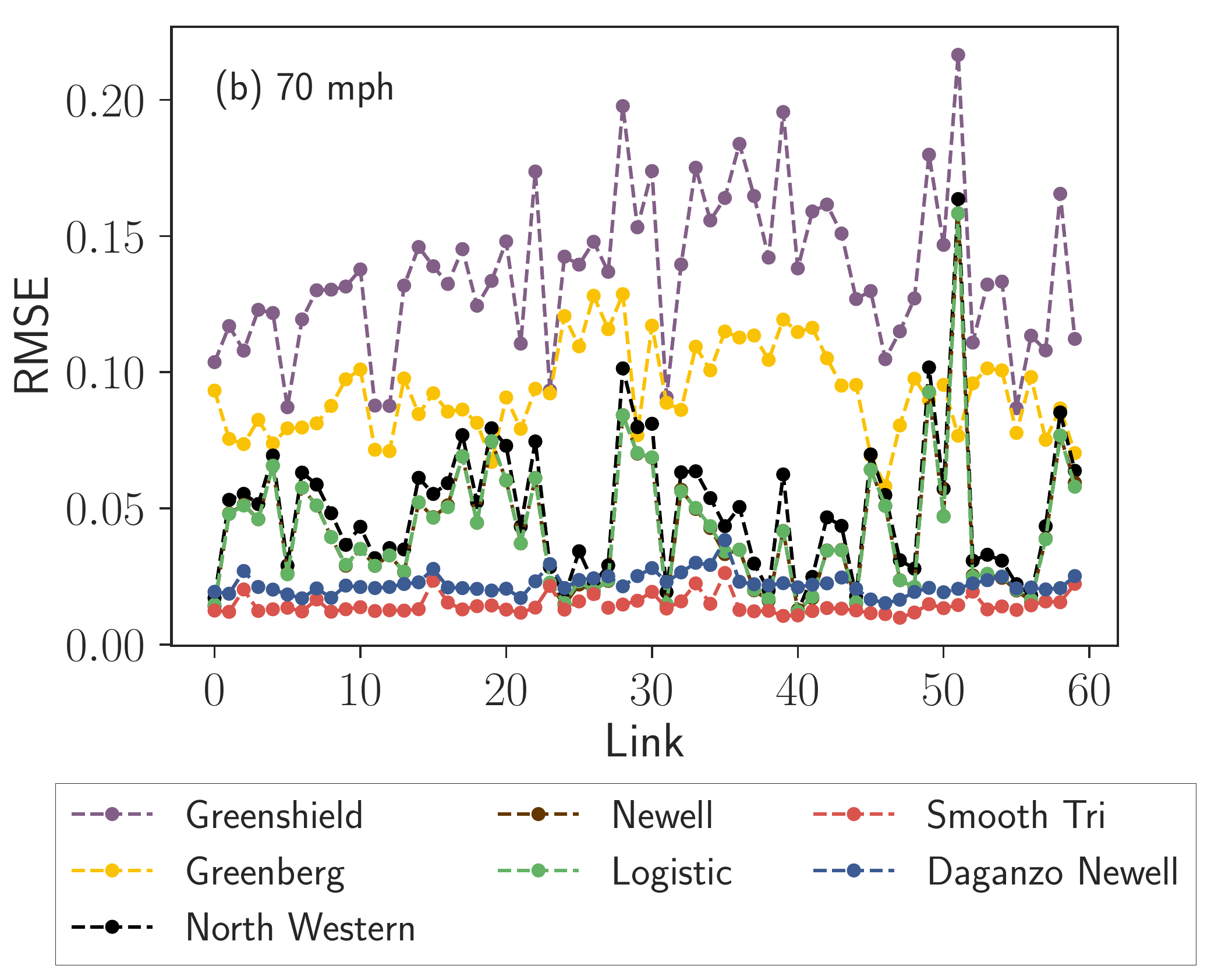}
	\end{subfigure}
	\caption[Optimal Diagram Analysis For All Links (Speed Limit Segmented)]{ Root mean square error results for all links considered. On top we fit only to data where 40mph speed limts are active. On the bottom we fit to data only when 70mph speed limits are active. 
	As in the non-segmented cases, we see the Daganzo Newell and Smooth Triangle results are generally superior across links, however it is far less clear in many cases, and in some they are indistinguishable from various other models. Note that on the 70mph diagram, the Newell and Logistic diagrams are almost indistinguishable as they have such similar RMSE values.
	}\label{fig:DiagramRMSE_Vs_SL}
\end{figure}

Contrasting Fig. \ref{fig:DiagramRMSE_Vs_Link} to Fig. \ref{fig:DiagramRMSE_Vs_SL}, we see that whilst for non-segmented data the triangular forms of diagram are clearly superior to all others, this becomes less clear when we segment by speed limit. 
All models perform similarly when 40mph speed limits are active, likely due to the large variation meaning we simply have outliers regardless of what functional form is used. 
In the 70mph case, we the smooth and discrete triangle diagrams are clearly superior to the Greenberg and Greenshields ones, however we see much more varied performance when considering the Logistic, North Western and Newell.
Generally, the difference in quality of fit between each diagrams is reduced when we segment by speed limit, and particularly at lower speed limits one could argue all diagrams are reasonably similar in description of the data, however at higher speed limits the separation becomes more visible.

\subsection{Spatial Consistency}\label{sec:LinkToLinkVariation}

Using this functional form, we can now question how spatially consistent the fundamental diagrams of the M25 are.
In particular, one may wonder if there exists one set of parameters that fits the M25, and all links vary around this, or if there are multiple subsets of parameters that best describe the traffic behaviour. 
To test this, we take the density-flow data for each diagram, again scaling it to ensure all links are independent of the number of lanes. 
We then fit the Daganzo-Newell fundamental diagram via least squares optimization. 
To ensure we do not get stuck in local minima, we start 100 such fits of each function, using random initial starting points and waiting for all to converge. 
Having done this for all links, we have a set of parameters for each link that best describe the fundamental diagram, and we then use clustering methods to discover natural patterns in the data.

As with all clustering methods, we re-scale the parameters to ensure our results are not dominated by a single input, and then using each links parameter set as an input, we cluster the fundamental diagrams, using Hierarchical Agglomerative Clustering (HAC) with Ward linkage \cite{WardLinkage} and euclidean distance.
Ward's linkage ensures the clustering result minimizes the variance observed in clusters.
Notice that we switch to HAC clustering rather than k-medoids here, as it allows us to visually see how our data separates into clusters, allowing an intuitive explanation of how many clusters are ultimately derived from the data.

\begin{figure*}[ht!]
	\centering
	\begin{subfigure}{0.48\textwidth}
		\includegraphics[width=\textwidth]{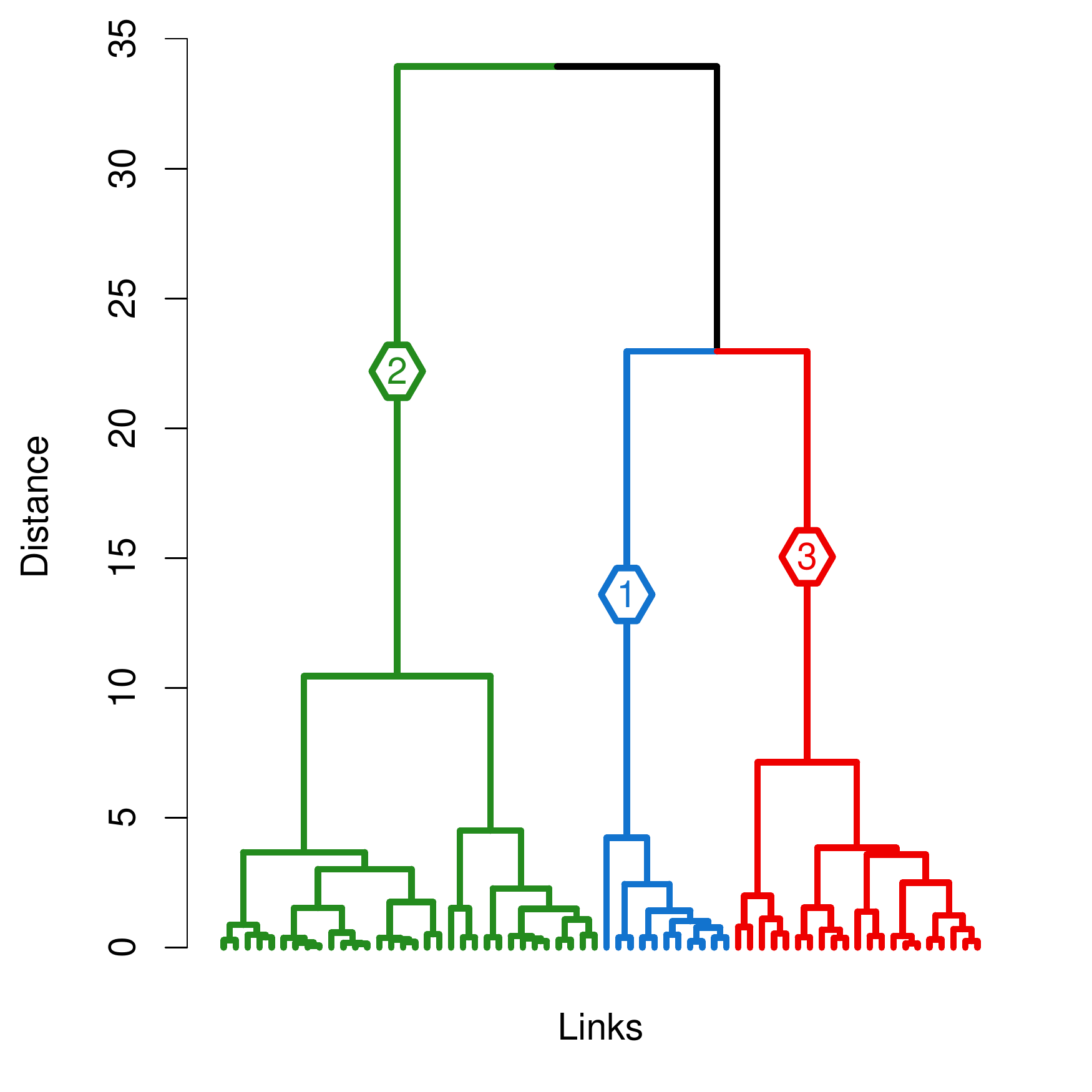}
	\end{subfigure}
	\begin{subfigure}{0.48\textwidth}
		\includegraphics[width=\textwidth]{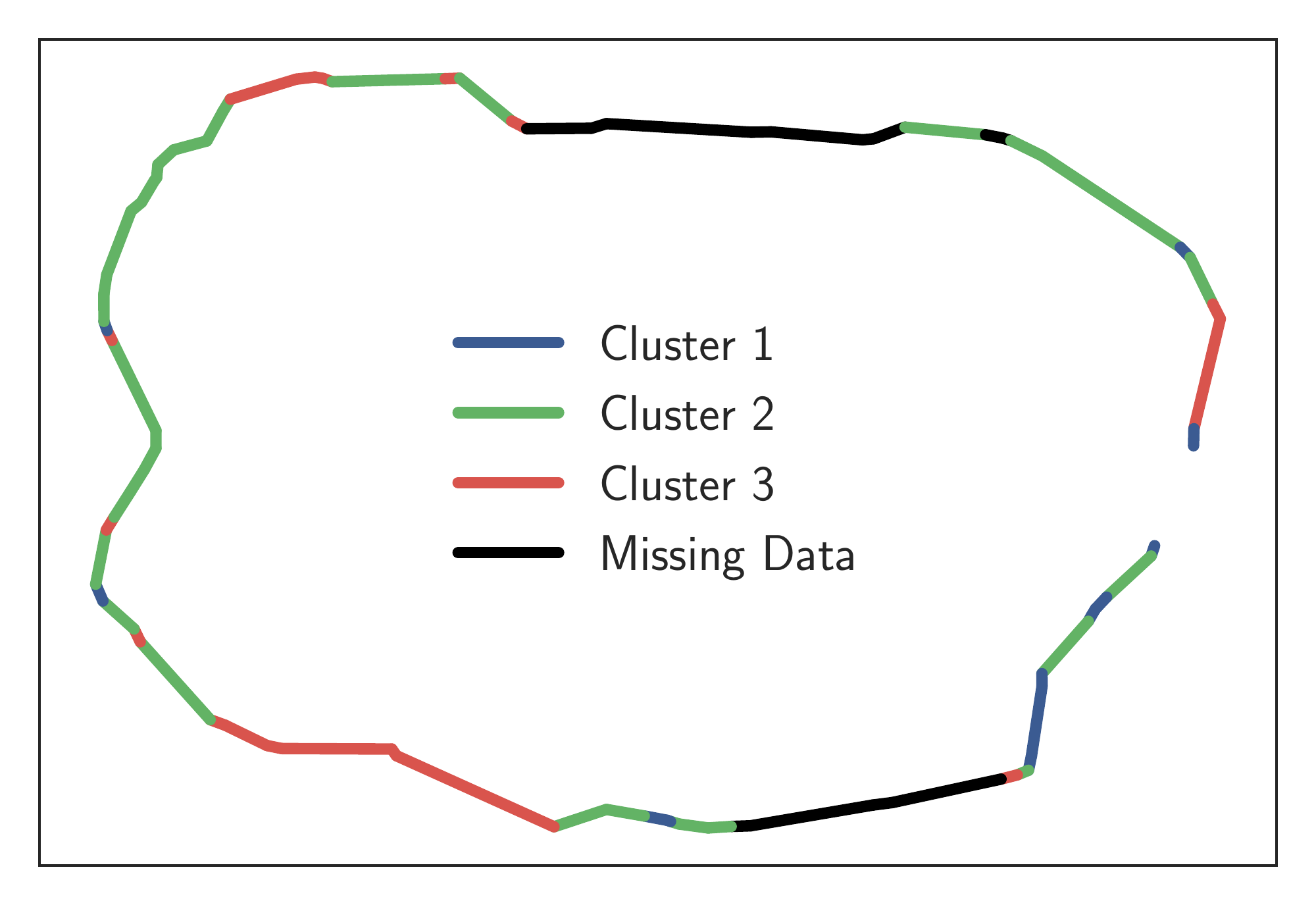}
	\end{subfigure}
	\caption[Link To Link Variation Clustering Results]{ Left: The dendrogram for the Daganzo-Newell fundamental diagram when fit to each of the links considered on the M25. Right: A plot of the M25, with links coloured by cluster membership. We see a clear separation into 3 clusters when inspecting the dendrogram, with a large cluster (green, 2) covering most of the west side and north-east of the M25. Cluster 1 (blue) is mainly concentrated in the south-east of the M25, although scattered around some other areas, and third cluster (red) mainly covers 3 separate areas in the south-west, north-west and east. }\label{fig:LinkToLinkClusteringResults_Dendogram}
\end{figure*}

After clustering, inspecting the dendrogram shows 3 natural parameter sets emerge in the data, suggesting 3 distinct fundamental diagrams when considering their optimal fit. 
We show the clustering dendrogram in Fig. \ref{fig:LinkToLinkClusteringResults_Dendogram} accompanied by a map showing the spatial locations on the M25.

Inspecting the map in Fig. \ref{fig:LinkToLinkClusteringResults_Dendogram}, we see evidence of spatial structure: the colours representing the three types of link identified by the clustering analysis form long chains rather than being scattered randomly. 
Additionally, we see two links in cluster 1 (blue) directly before and after the M25 meets the Dartford tunnel and A282, where one would expect different behaviour to say, a link in the center of the M25.

\begin{figure*}[ht!]
	\centering
	\begin{subfigure}{0.32\textwidth}
		\includegraphics[width=\textwidth]{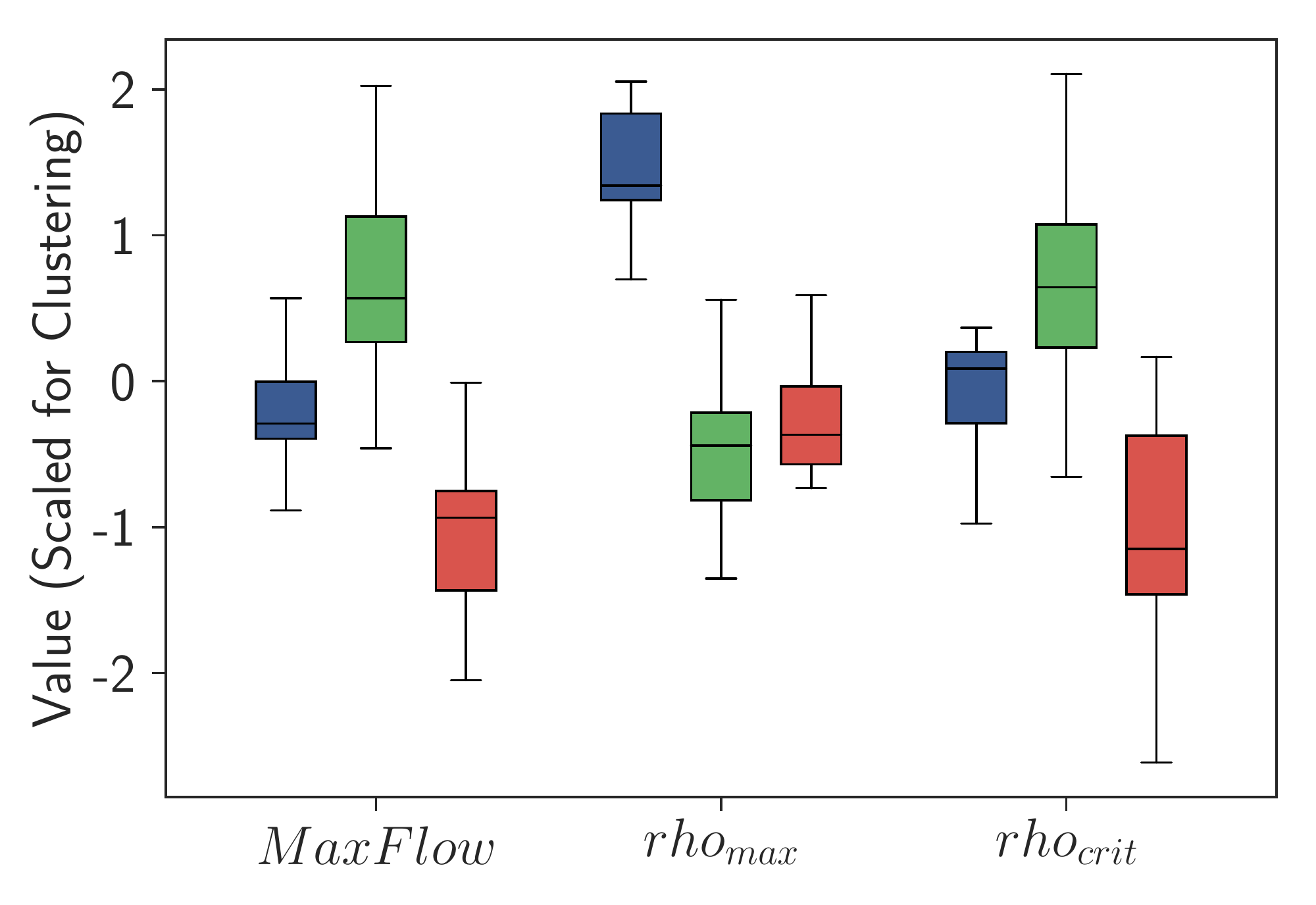}
	\end{subfigure}
	\begin{subfigure}{0.32\textwidth}
		\includegraphics[width=\textwidth]{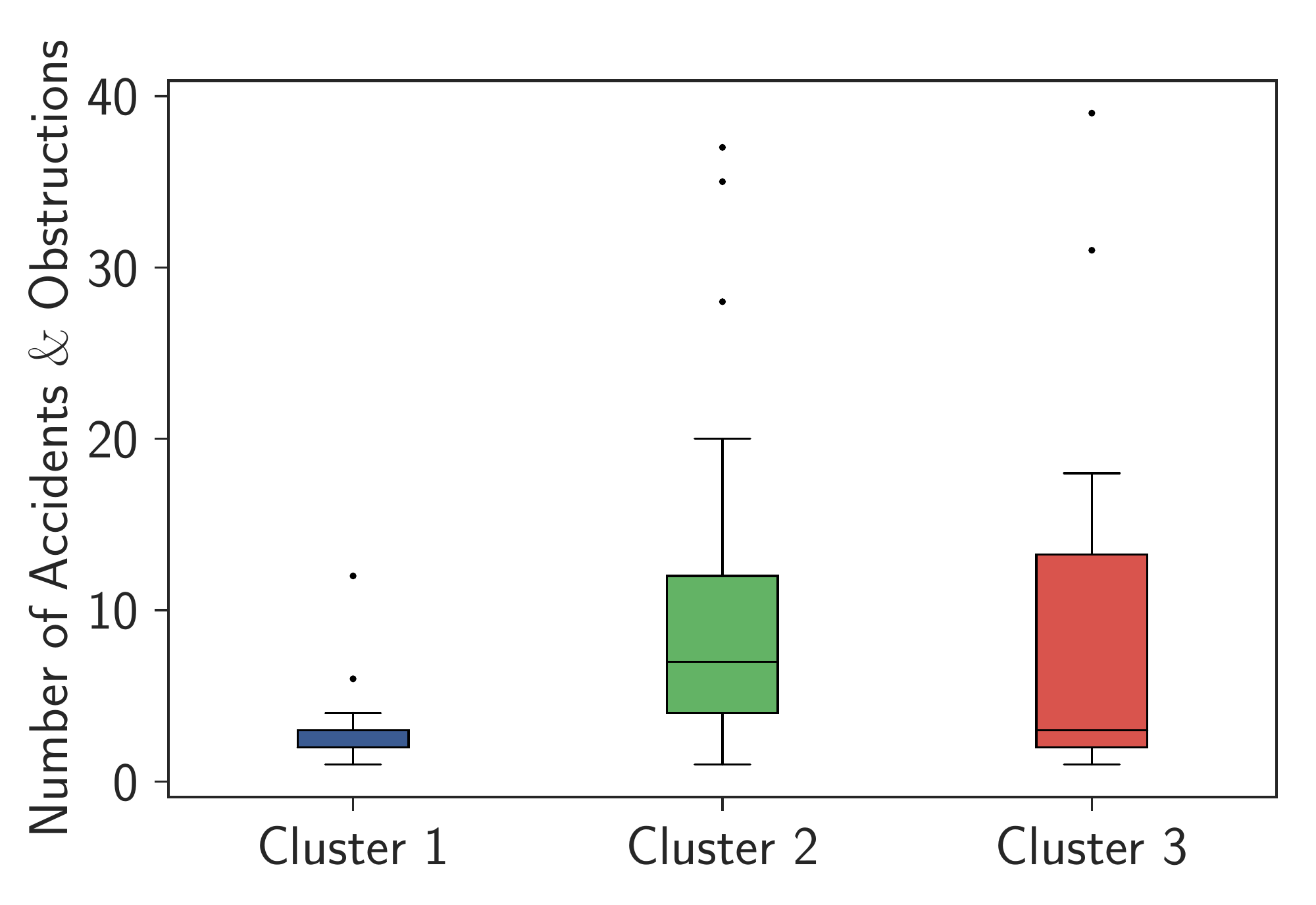}
	\end{subfigure}
	\begin{subfigure}{0.32\textwidth}
		\includegraphics[width=\textwidth]{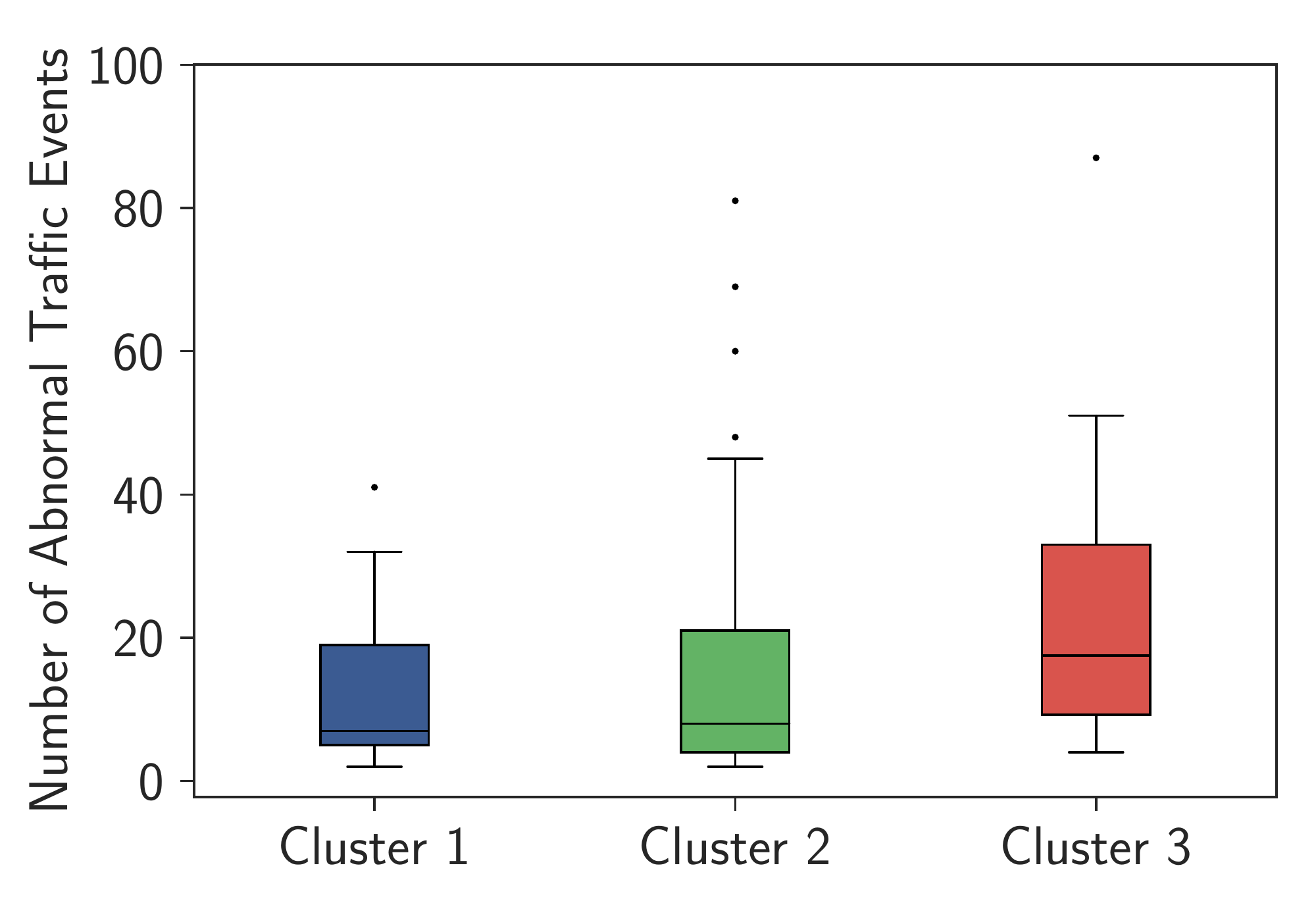}
	\end{subfigure}
	\caption[Parameter Values for Identified Clusters]{ Left: Box plot of the parameter values in each cluster observed during Daganzo-Newell diagram clustering. Center: Box plot of the number of accidents and obstructions on each link in each cluster. Right: Box plot of the number of abnormal traffic events on each link in each cluster. From the parameters box plot,  we see a far larger $\rho_{max}$ values in cluster 1 than any other cluster. Similarly, there is clear separation of the maximum flow parameters in cluster 2 (green) compared to the other clusters. Cluster 3 appears to have lower maximum flow and $\rho_{crit}$ values than any other cluster. }\label{fig:FD_Cluster_Param_Hists}
\end{figure*}

Having identified 3 clear clusters of parameters and spatial consistency, we now consider why these groups exist.
As we have clustered the diagram parameters, we can directly relate these to link behaviour.
Additionally, we can consider the influence of events by counting the occurrences on each link in each cluster.
Results for each parameter set, the number of accidents and obstruction events, and the number of abnormal traffic events are given in Fig. \ref{fig:FD_Cluster_Param_Hists}.

Inspecting Fig. \ref{fig:FD_Cluster_Param_Hists}, we see clear differences in the parameter values within each cluster. 
Note first that, as all density and flow values were scaled before fitting, these fits should be independent on the number of lanes a link has as well as the length. 
We see cluster 1 (blue) has by far the highest maximum density observed of any of the links. 
A high maximum density would imply that, during the breakdown phase of flow, higher densities do not cause as severe decreases in flow than in other links. 
As a result, these links may be those that are more resistant to severe congestion when breakdown does occur. 

Cluster 2 (green) has significantly higher maximum flow parameters than any others, as well as critical densities. 
From this, we would infer that these links are typically reaching a high capacity more often than links in other clusters, and flow breakdown occurs at higher densities than other clusters links. 
In a spatial sense, these links are mostly located on the west and north of the M25.

In contrast, cluster 3 (red) has far lower maximum flow parameters than either other cluster, as well as lower critical densities. 
In a real-sense, this means we would expect these links to reach high capacities less often than links in other clusters, and have flow breakdown occur at earlier densities than normal.
The majority of these links can be found in the south-west of the M25, as well as in the north-west and east. 
The eastern links are before the Dartford crossing, although note the single links directly before the crossing are both members of cluster 1 (blue). The south-western links are roughly near Crawley and Guildford.

It is also interesting to note that cluster 1 (blue) has far fewer accidents and obstructions observed than clusters 2 (green) and 3 (red), but the amount of abnormal traffic events is comparable across all clusters.
Accidents and obstructions may lead to significant breakdown in the density-flow relationship, however as abnormal traffic events are defined by some arbitrary threshold, they may not necessarily describe the same severity of breakdown as an accident would. 
If so, this would explain the large $\rho_{max}$ parameters observed in this cluster.

\begin{figure}[ht!]
	\centering
	\begin{subfigure}{\linewidth}
		\includegraphics[width=\textwidth]{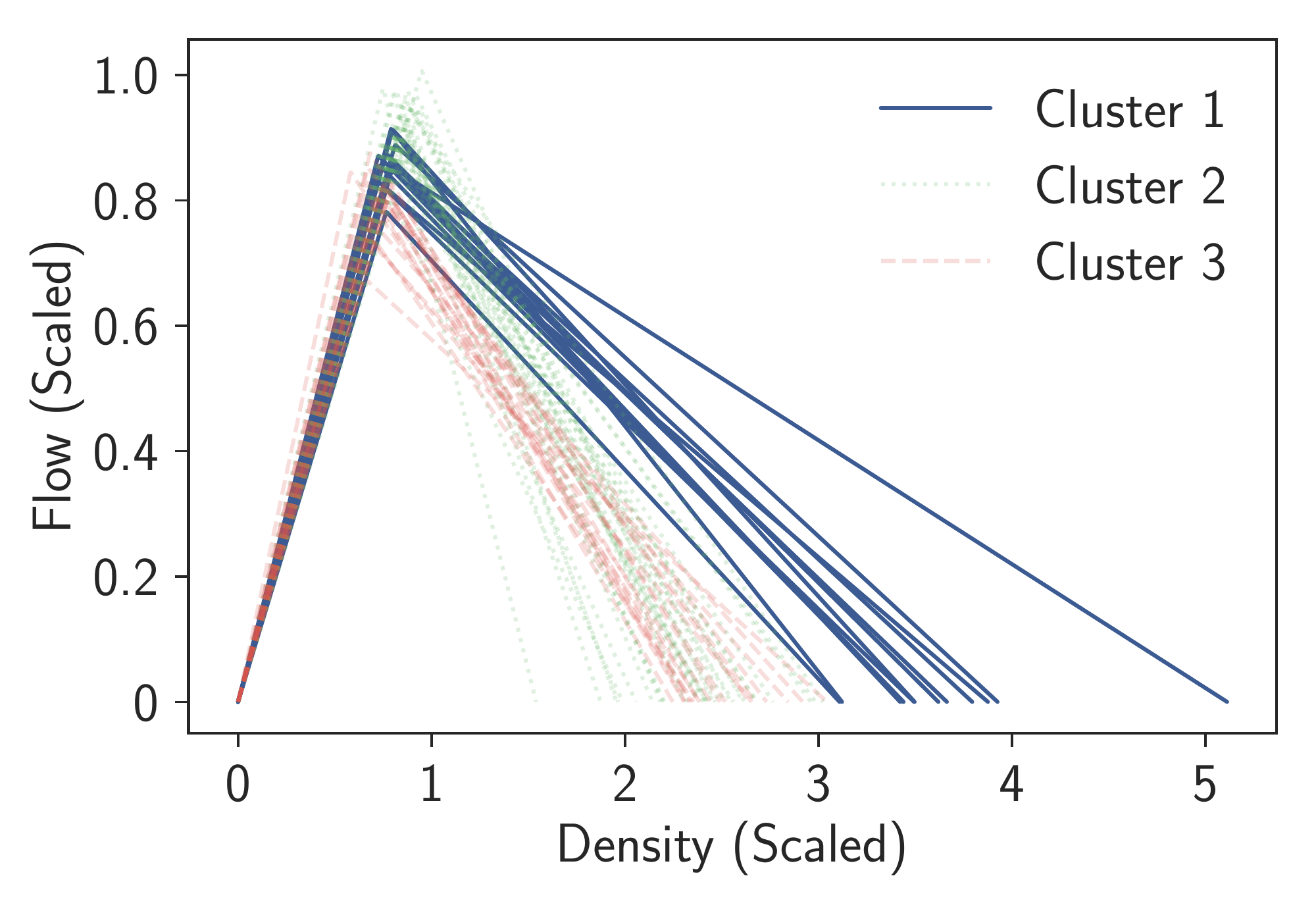}
	\end{subfigure}
	\begin{subfigure}{\linewidth}
		\includegraphics[width=\textwidth]{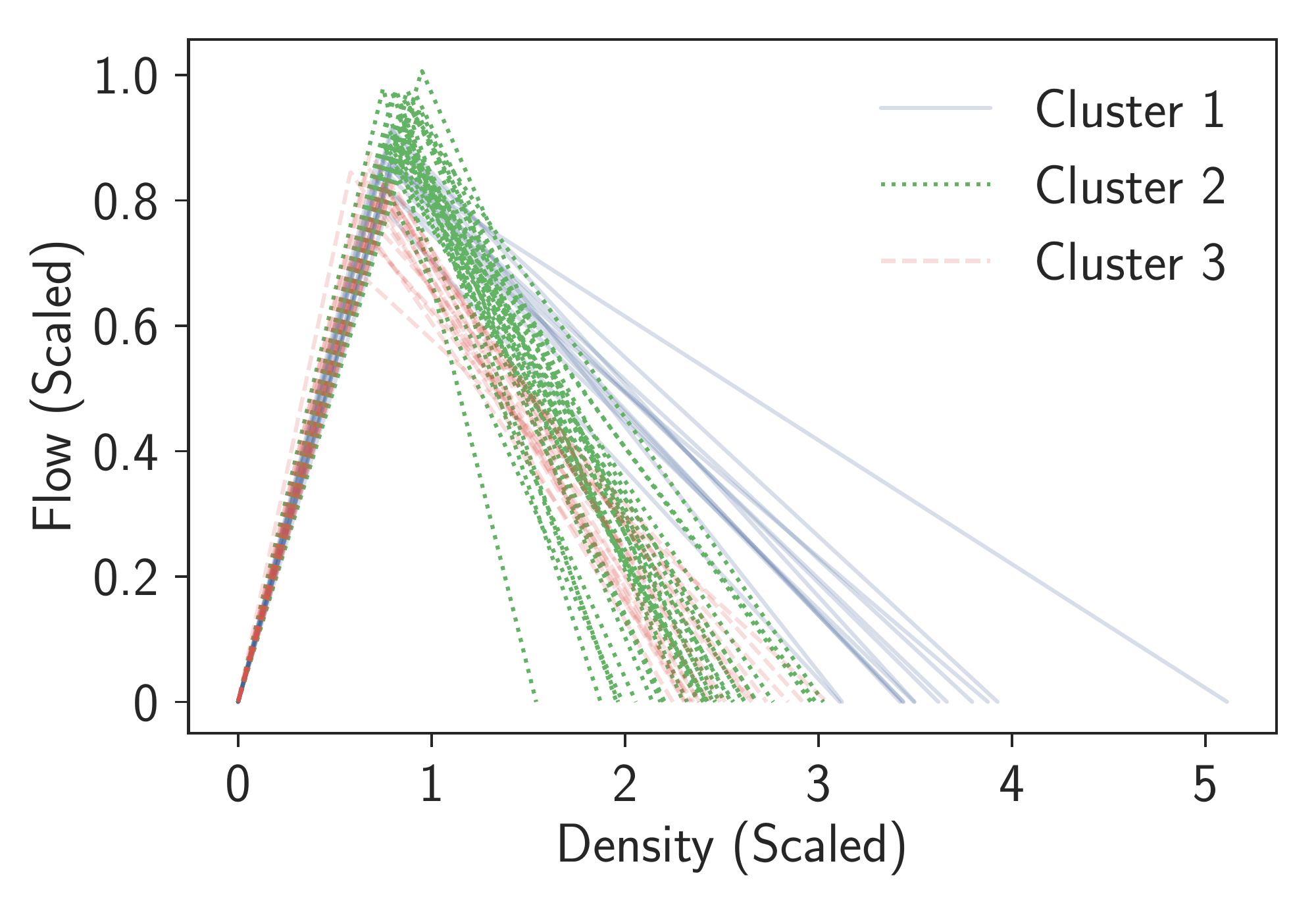}
	\end{subfigure}
	\begin{subfigure}{\linewidth}
		\includegraphics[width=\textwidth]{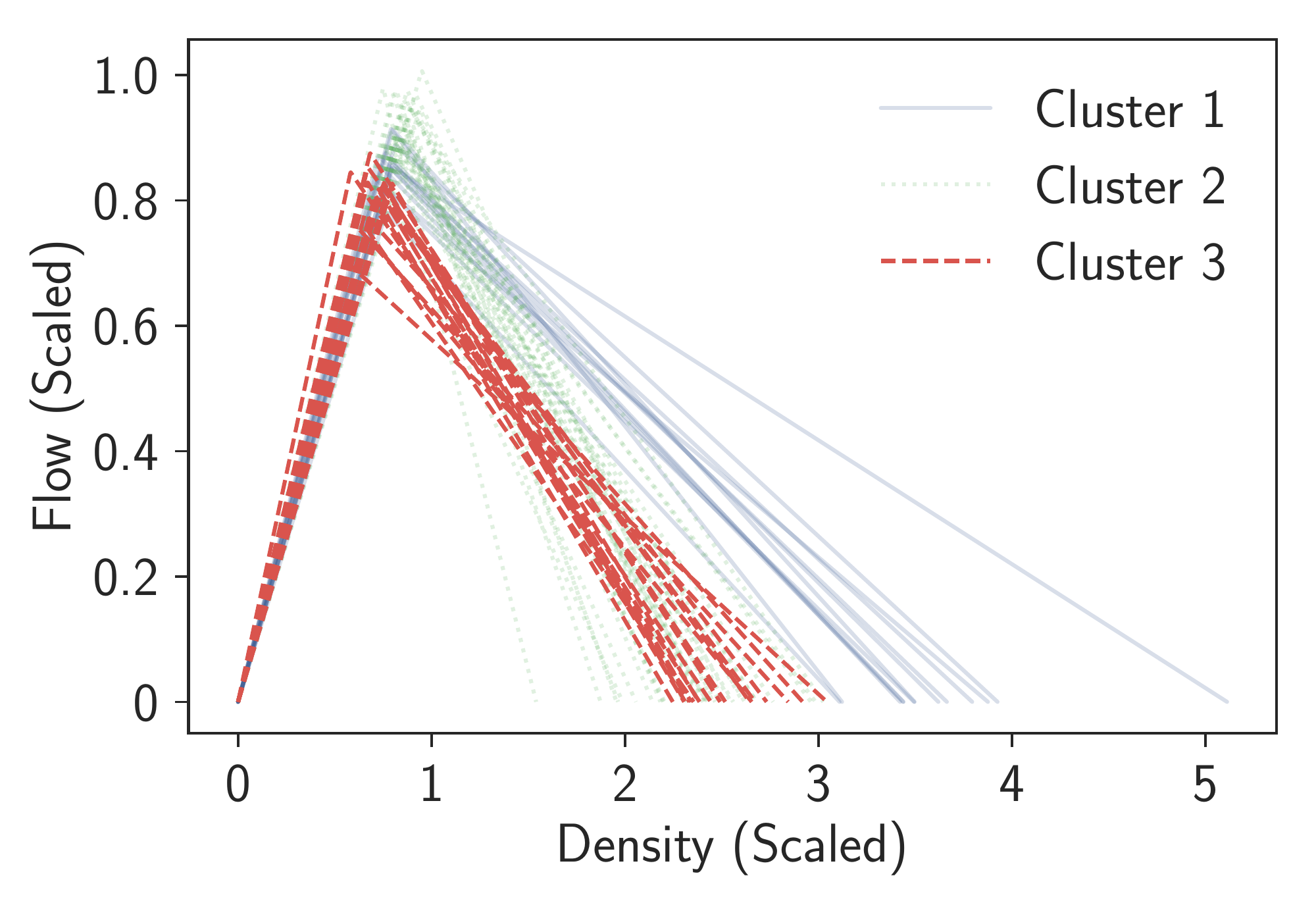}
	\end{subfigure}
	\caption[Plot of all fits]{ Plots of all optimal fundamental diagrams, coloured and line-styled by cluster membership. The top diagram highlights cluster 1 (blue), the center cluster 2 (green) and the bottom cluster 3 (red). Here we see how the differences in parameters observed in Fig. \ref{fig:FD_Cluster_Param_Hists} relate directly to the functional fits of the diagrams. }\label{fig:FD_Cluster_AllDiags}
\end{figure}

A final, and potentially most useful visualization of the cluster differences is simply to plot all of the diagrams for a direct comparison. 
Such a plot is given in Fig. \ref{fig:FD_Cluster_AllDiags}, where we plot each cluster as coloured in the previous plots, but include all other diagrams faded for reference.

Considering Fig. \ref{fig:FD_Cluster_AllDiags}, we see cluster 1 (blue) clearly has the most elongated breakdown phase (large $\rho_{max}$), where as cluster 2 has all of the highest peaks and some of the most severe breakdowns (low $\rho_{max}$). 
Cluster 3 (red) typically has flow breakdown occur earlier than any cluster (low $\rho_{crit}$), shown by having the lowest density maxima, but it is typically neither the most or least extreme (intermediate $\rho_{max}$ values).

\section{Discussion \& Conclusions}

In this paper, we have taken a large dataset from a smart motorway in the UK and considered ways to measure the variation found in fundamental diagrams of traffic flow.
We have approached this by looking at the variation when segmented by speed limits active, as well as the spatial dependence of the parameter sets of the most appropriate identified functional forms to represent the data.
Our results not only offer a better understanding of the dynamics of traffic behaviour as viewed from the standpoint of a fundamental diagram, but they also highlight how useful different interpretations may be, rather than simply visual checks of scatter plots. 

We found two modes exist in typical fundamental diagrams, a high and low density mode, and quantified the spread of data around each mode.
We found heavy tail distributions of deviations from these modes, and showed how the range of these decayed as we moved from a stable (70mph speed limit) to a less stable (40mph speed limit) case. 
Comparing various links, we found 3 clusters of parameter sets exist in fundamental diagrams, and investigated why this may be so. 
Our investigation of the parameters in each cluster showed spatially where high capacity, relative to other links is most often achieved on the M25, as well as where the earliest breakdowns in flow occurred. 
Finally, we considered the frequency of events that would impact flow on links, relating these to both the clustering found and also spatially to the M25. 

In the future, our results may be useful when considering transportation planning and road management, particularly on where to focus improvements in infrastructure to the M25 or how to understand and quantify the usage of variable speed limits.  
Further work in this area could address a limitation of our dataset, the lack of measures of severity of events. 
Currently, an event could simply be debris blocking a single lane on a link, or an overturned vehicle blocking multiple lanes. 
Additional data to consider the severity of each event in each cluster, as well as some quantitative measure of severity, could provide further insight into this 3 fundamental diagram behaviour.

\appendices

\section{Functional Forms of Diagrams}\label{appendix:FuncForms}

Table \ref{table:FundDiags} lists a number of common functional forms of fundamental diagrams. 

\begin{table*}[ht!]
\centering
	\scriptsize
	\begin{tabularx}{0.873\linewidth}{ |C{1.6cm}|C{8.2cm}|C{3.3cm}| }
	\hline
	Source & Flux Function & Parameters \tabularnewline
	\hline
	Greenshield  \cite{GreenshieldsFD} & $f^{FD}_{Greenshield}(\rho) = \rho v_{free} \left( 1 - \frac{\rho}{\rho_{max}} \right)$     & \makecell{$v_{free}$ - Free flow velocity  \\ $\rho_{max}$ - Maximum road density }  \tabularnewline
	\hline
	Greenberg \cite{GreenbergFD}       & $f^{FD}_{Greenberg}(\rho) = \rho v_{capacity}\ln\left( \frac{\rho_{max}}{\rho} \right)$   & $v_{capacity}$ - Velocity when road is at full capacity \tabularnewline  
	\hline 
	Northwestern \cite{DrakeFD}        & $f^{FD}_{NW}(\rho) = \rho v_{free}e^{ -\frac{1}{2}\left( \frac{\rho}{\rho_{crit}} \right)^2 }$                                                 & $\rho_{crit}$ - Density when road is at full crit \tabularnewline 
	\hline
	Newell \cite{NewellFD}             & $f^{FD}_{Newell}(\rho) = \rho v_{free}\left( 1 - e^{\left[ -\frac{c_1}{v_{free}}\left( \frac{1}{\rho} - \frac{1}{\rho_{max}} \right) \right]} \right)$ &  $c_1$ - Decay calibration coefficient \tabularnewline 
	\hline
	Logistic Model \cite{LogisticFD}   & $f^{FD}_{Logistic}(\rho) = \rho \frac{v_{free}}{1 + e^{\left( \frac{\rho - \rho_{crit}}{c_2} \right)}}$ & $c_2$ - Decay calibration coefficient \tabularnewline 
	\hline
	Daganzo-Newell \cite{DaganzoCTM, NewellCTM} & 
														\vspace{-3mm}{\begin{equation*}
														 f^{FD}_{tri}(\rho) = \begin{cases} 
														      MaxFlow\frac{\rho}{\rho_{crit}} & \mbox{for } \, \, 0 \leq \rho \leq \rho_{crit} \\
														      MaxFlow\frac{\rho_{max}-\rho}{\rho_{max}-\rho_{crit}} & \mbox{for } \, \, \rho_{crit} < \rho \leq \rho_{max}
														    \end{cases}
														 \end{equation*} } 
												    & $MaxFlow$ - The maximal flow possible on the road \tabularnewline
	\hline
	Continuous Triangle \cite{FanHertyDataFitGARZ} & 
														\vspace{-3mm}{\begin{equation*}
														\begin{split}
														 &f^{FD}_{\alpha, \lambda, p}(\rho) = \alpha \left( a + (b - a)\frac{\rho}{\rho_{max}} - \sqrt{1 + y^2} \right) \\
														 &a = \sqrt{1+ (\lambda p)^2}, \, \, \, b = \sqrt{1 + (\lambda(1-p))^2}, \, \, \, y = \lambda \left( \frac{\rho}{\rho_{max}} - p \right)
														\end{split}
														 \end{equation*} } 
												    & \makecell{$\alpha$ - Scaling parameter  \\ 
															    $\lambda$ - Turning point parameter \\
															    $p$ - Curvature parameter } \tabularnewline
    \hline
    \end{tabularx}
	\caption[Typical Fundamental Diagram Functions]{Typical fundamental diagrams found in the literature.}\label{table:FundDiags}
\end{table*}

\section*{Acknowledgments}

The authors are grateful to Dr. Steve Hilditch and Thales UK for sharing expertise on the operation of NTIS and UK transportation systems more generally.  We also thank Ayman Boustati and Alvaro Cabrejas Egea  for help with data acquisition and pre-processing. 


\bibliographystyle{plain}
\bibliography{fund_diag_variation}

\end{document}